\documentclass[twocolumn,usenatbib]{mnras}

\usepackage{newtxtext,newtxmath}
\usepackage{esvect}
\usepackage[font=small,skip=0pt]{caption}

\usepackage{comment}

\usepackage{graphicx}
\usepackage{appendix}
\usepackage{float}
\title[Torques on FIRE]{Angular momentum transfer in cosmological simulations of Milky Way-mass discs}

\author[C. Trapp et al.]{
Cameron W. Trapp$^{1}$,\thanks{E-mail: ctrapp@ucsd.edu}
Du\v{s}an Kere\v{s}$^{1}$,
Philip F. Hopkins$^{2}$,
Claude-Andr\'e Faucher-Gigu\`ere$^{3}$,
\newauthor
and Norman Murray$^{4}$
\\
$^{1}$Department of Astronomy and Astrophysics, University of California San Diego, 9500 Gilman Dr, La Jolla 92093, USA\\
$^{2}$TAPIR, Mailcode 350-17, California Institute of Technology, Pasadena, CA 91125, USA\\
$^{3}$Department of Physics and Astronomy and CIERA, Northwestern University, 1800 Sherman Ave, Evanston, IL 60201\\
$^{4}$Canadian Institute for Theoretical Astrophysics, 60 St. George Street, University of Toronto, ON M5S 3H8, Canada\\
}

\date{Accepted XXX. Received YYY; in original form ZZZ}

\pubyear{2023}

\newcommand{\tgal}{$t_{\rm orbit}$ }
\newcommand{\rgal}{$R_{\rm DLA}$ }
\newcommand{\rgalStop}{$R_{\rm DLA}$}
\newcommand{\hgal}{$h_{\rm total}$}

\newcommand{\MFunit}{M$_{\odot}$ yr$^{-1}$}

\begin{document}
\label{firstpage}

\maketitle

\begin{abstract}

  Fueling star formation in large, discy galaxies requires a continuous supply of gas accreting into star-forming regions. Previously, we characterized this accretion in 4 Milky Way mass galaxies ($M_{\rm halo}\sim10^{12}M_{\odot}$) in the FIRE-2 cosmological zoom-in simulations. At $z\sim0$, we found that gas within the inner circumgalactic medium (iCGM) approaches the disc with comparable angular momentum (AM) to the disc edge, joining in the outer half of the gaseous disc. Within the disc, gas moves inward at velocities of $\sim$1-5~km~s$^{-1}$ while fully rotationally supported. In this study, we analyze the torques that drive these flows. In all cases studied, we find that the torques in discs enable gas accreted near the disc edge to transport inwards and fuel star formation in the central few kpc. The primary sources of torque come from gravity, hydrodynamical forces, and the sub-grid $P dV$ work done by supernova (SNe) remnants interacting with gas on $\lesssim$10 pc scales. These SNe remnant interactions induce negative torques within the inner disc and positive torques in the outer disc.  The gas-gas gravitational, hydro, and "feedback" torques transfer AM outward to where accreting gas joins the disc, playing an important role in driving inflows and regulating disc structure. Gravitational torques from stars and dark matter provide an AM sink within the innermost regions of the disc and iCGM, respectively. Feedback torques are dominant within the disc, while gravitational and hydrodynamical torques have similar significance depending on the system/region. Torques from viscous shearing, magnetic forces, stellar winds, and radiative transfer are less significant.

\end{abstract}


\graphicspath{{./}{Figures/}}

\section{Introduction}

The mechanism by which gas is able to fuel active star-forming regions in massive disc galaxies has been a long-standing problem in our understanding of galaxy formation. These galaxies, including our own Milky Way (MW), show relatively stable star formation rates (SFRs) over the past few Gyr \citep{binney00}. In order to sustain these SFRs, an active supply of gas is needed, as existing H$_{2}$ reservoirs within the present-day interstellar medium (ISM) would be depleted within $\sim$1-2 Gyr \citep{tacconi18,saintonge17}. Even extended HI reservoirs of large discy galaxies, which can condense to H$_{2}$, would still deplete within $\sim$2 Gyrs \citep{kennicutt98} without new inflow. Gas recycling via stellar mass loss \citep{leitner11} can help refill these reservoirs, but a continuous supply of gas is still necessary. Local observations, such as the G-Dwarf problem (i.e. the relative scarcity of low metallicity stars in the solar vicinity does not match predictions from simple galactic chemical evolution models due to continuing accretion of gas enabling star formation at later times, where average metallicity will be higher) \citep{bergh62,schmidt63,sommer-larsen91,worthey96,haywood19}, additionally motivate the need for continuous accretion of gas from the circumgalactic medium. 

Historically, one of the most direct ways to search for this accretion is through observation of Intermediate- and High-Velocity Clouds (IVCs and HVCs). IVCs and HVCs are gaseous clouds with strong kinematic deviations from galactic rotation, so this classification selects gas that is not yet part of the rotationally supported disc. These deviations are typically within 40-70 km/s for IVCs and above 70-90 km/s for HVCs \citep{Rohser18}. Observations of HVCs around the MW have shown total gas accretion rates of around 0.4 $\rm M_{\odot}$/yr, which is not enough to fully support observed SFRs of $\sim$2-3$\rm M_{\odot}$/yr \citep{putman12}. 

In a previous study, we characterized gas accretion and radial transfer through galactic discs \citep{trapp22} in the FIRE-2 simulations. We found that accretion from gas that corresponds to HVCs is indeed only a small fraction of the total needed to fully sustain SFRs of star-forming disc galaxies, implying that the majority of gas inflow is largely co-rotating with the disc. 

In the CGM, there is also indirect observational support for this disc accretion scenario from larger scale gas flows, where cold/warm absorbers in halos of low redshift galaxies show co-rotation with the disc, potentially mapping such infalling gas \citep{martin12,bielby17,peroux17,diamond16,muzahid16}. Studies at higher redshifts utilizing background quasars to probe disc outskirts have also shown kinematic evidence for co-rotating structures out to 30-60 kpc \citep{barcons95,bouche13,zabl19}.  See \cite{tumlinson17} and \cite{cf&oh_23} for thorough reviews on relevant CGM processes.

How gas accretes over cosmic time from the inter-galactic medium to galactic regions has been studied extensively in hydrodynamic galaxy formation simulations. Generally, these simulations show that star formation is largely supply-driven, with typical accretion rates on the order of galactic star formation rates \citep[e.g.][]{keres05}. Simulations have found that gas can accrete in "cold-mode", which dominates at early times \citep{keres05, dekel06, ocvirk08, brooks09, keres09a,faucher11,faucher11:SmallCoveringFactorOfColdAccretionStreams, vandevoort11a,stern20b}, and "hot-mode", which dominates at later times and/or in more massive halos. In cold-mode accretion, gas accretes along filamentary streams and does not shock to the virial temperature in the outer halo.
The Milky Way is currently expected to be in the "hot mode" regime, but close to the mass where the transition between these two cases occurs \citep{keres05,dekel06, ocvirk08, vandevoort11b, nelson13,stern19:CoolingFlowSolutionsForCGM,stern20a,hafen20:FatesOfCGMInFIRE,hafen22}. Details of this transition depend on feedback and nonthermal pressure \citep{ji21:virial_shocks_in_cr_halos}.

Simulations suggest that high-redshift gas accretion can co-rotate with the disc near the edge of the stellar disc \citep{keres05, Danovich15, stewart17}; however, the relative burstiness of the stellar feedback at these times leads to chaotic discs with large velocity dispersions and strong time variations in infalling gas \citep[e.g.][]{muratov15,gurvich23:RapidDiscSettling}. At later times, infall tends to be steadier, while outflows are weaker, enabling the formation of a coherent disc \citep[e.g.][]{kassin12, muratov15,stern20b,stern24:HotRotatingCGMInflows}. It has been shown that accreting cold gas at these times is largely corotating \citep{keres09b, stewart11b, ho19}. Given the large angular momentum of overall halo gas, which is comparable to but higher than that of the dark matter halo \citep{elbadry18,romeo23}, accreting gas can settle into rotational support in the disc outskirts regardless of its temperature history, consistent with the standard picture of inside-out disc growth \citep{fall80}.\footnote{The buildup of a deep central potential, details of the rotating cooling flows, and lack of bursty star formation all likely contribute to the formation of the late time discs \citep{hopkins23}.}

Previous work based on the FIRE-2 simulations has shown that a continuous supply of gas radially transporting from the inner circum-galactic medium (iCGM) allows for observed SFRs to be maintained \citep{trapp22, hafen22}. In brief, we found that these galaxies show a sharp gaseous disc edge, with HI column densities dropping 4-5 orders of magnitude within $\lesssim$10 kpc. Gas joins the disc from the CGM at time/azimuthally-averaged radial speeds of $\sim$10-20 km/s along trajectories at small, but nonzero angles ($\sim$15\textdegree) above the disc plane. This gas joins in the outer half of the gaseous disc, predominantly a few kpc inside of the gaseous disc edge. Once within the disc, gas slows down to average radial speeds of a few km/s ($\sim$1-5 km/s). Averaged over time, these flows roughly correspond to the SFRs of these galaxies ($\sim$ 2-3 \MFunit).  This overall trend has intrinsically large variations in time and space. The instantaneous velocity of the gas is higher, with a typical dispersion of the cold gas around 10 km/s \citep{chan22} and deviations up to $\sim$ 40-60 km/s, showing strong oscillatory behavior both spatially and temporally.

Gas outside the disc is largely co-rotating with the disc edge, but it is not fully rotationally supported. This gas will therefore radially transport inwards, without the need for additional torque. However, it still loses a small amount of angular momentum as it moves inwards and gains angular momentum just before settling into full rotational support. Once inside the disc, radially transporting gas is strongly co-rotating and is on average fully rotationally supported at all radii. Given the average radial transport speeds, it must therefore be losing angular momentum. Identifying the sources of torque responsible for these changes in angular momentum will give insight into how these galaxies evolve and what ultimately governs their growth. Note that caution is needed when interpreting changes in angular momentum as inflow/outflow. This is expected to hold well for homogenous, well-mixed gas in slowly decaying circular orbits and should therefore hold for gas in the late-time discs considered in this study. It will not generally apply, particularly for gas that is already strongly inflowing/outflowing.

One source of torque is the gravitational forces on gas from stars, dark matter, and gas self-gravitation, which can provide an important source of non-local angular momentum transport in the galactic disc. Gravitational forces can transfer angular momentum from more interior regions outward to the outer disc or halo. Resonances in spiral-like structure and dynamical friction \citep{lyndenBell72, tremaine84} have been shown to transfer angular momentum outward. Bar-like structures, in particular, have been shown to act as a sink for angular momentum in the inner disc, ultimately transporting it to the dark matter halo \citep{hernquist92,weinberg02,weinberg07a,weinberg07b,ceverino07,petersen19}. The direct effect of these gravitational torques has generally been found to be very inefficient, unless close to co-rotation.

Instead, gravitational torques can torque gas into supersonic local relative motion, which induces strong shocks. This generates dissipation, which can lead to more significant net changes in angular momentum, as well as generating strong density enhancements and star formation \citep{quataert11}. While the direct effects from gravitational torques may not be significant enough to drive observed motions, the effects from the hydrodynamical shocks and star formation they induce may be more significant on these scales.

Hydrodynamical forces in general can provide a mechanism for local transport of angular momentum between neighboring gas mass elements. Pressure gradients within the gaseous disc, viscous shearing of gas due to differential rotation, and other turbulent hydrodynamical effects have been shown to transfer angular momentum on various scales in Keplerian discs \citep{shakura73,lodato04}. Likewise, magnetic forces can also play a significant role in shaping and aligning accretion flows onto Keplerian discs \citep{balbus91,terquem96,varniere02,wang22}. It is expected that hydrodynamical shocks induced by gravitational torques will play an important role \citep{quataert10,quataert11}, but it remains unclear how efficient these sources of torque will be in the context of a galactic disc.

In previous studies of the FIRE simulations, it was found that the inclusion of effects such as magnetic fields and viscosity, among others, have limited effects on the growth and evolution of galaxies and their star formation rates (SFRs) \citep{su17,hopkins20}. Instead, the strength of stellar feedback had the most salient effect on the evolution and growth of galactic discs. Analyzing the torques that arise from stellar winds and supernovae, as well as radiative forces from stellar populations, is therefore necessary for a complete understanding of how the angular momentum of gas evolves. Additionally, \cite{prieto16-multiscaleMassTransport} and \cite{prieto17-FeedbackEffectsOnMassTransportAtHighZ} found that torques from pressure, likely driven by feedback effects from SNe, were quite significant in driving mass transfer on disc scales in the context of feeding central supermassive black holes up to redshift z$\sim$6. It is not clear if similar effects will be significant in driving gas flows in low redshift, MW mass discs.
  
In this study, we investigate the relative contributions of various sources of torque on gas within 4  Milky Way mass discs in the FIRE-2 (Feedback In Realistic Environments)\footnote{http://fire.northwestern.edu} cosmological zoom-in simulations, which in addition to explicit stellar feedback models, incorporate magnetohydrodynamics (MHD) and cosmic ray physics (CR+; \citealt{chan19,hopkins20}). Specifically, we analyze the following sources: gravitational torques from gas, stars, and dark matter; MHD torques from pressure gradients, viscous shearing, non-continuum hydrodynamical terms, and magnetic fields; direct momentum injection from stellar feedback, including supernovae and stellar winds; and finally torques from the forces arising from radiative transfer. We additionally break down the pressure torques, investigating the relative contributions of thermal pressure, cosmic ray pressure, and magnetic pressure separately.

The paper is outlined as follows.  In Section~\ref{sec:methodology} we give an overview of the simulations and how the sources of torques were measured. In Section~\ref{sec:results} we go over the quantitative and qualitative behavior of the torques. In Section~\ref{sec:discussion} we discuss which torques are dominant in different regions and how angular momentum is being transferred within the disc. We additionally discuss the numerical implementations of SNe in the FIRE-2 simulations, how these torques relate to the measured radial velocities in these simulations, and briefly discuss observational implications for these results.

\section{Methodology} \label{sec:methodology}

\subsection{Galaxy Sample} \label{sec:methodology-galaxySample}
\begin{figure}
    \center{\includegraphics[width=.5\textwidth]
 	       {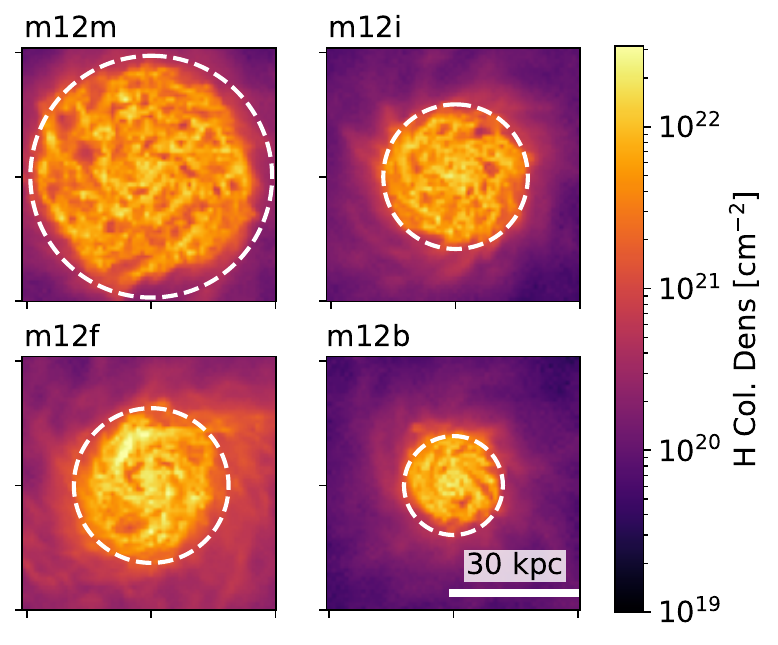}}
	  \caption{\label{fig:facePlot} Total hydrogen column density along the line of sight for the 4 galaxies in our sample at redshift z=0. Pixels in this and subsequent projection maps are 500x500 pc$^{2}$. White dashed lines show the radius where the neutral hydrogen column density drops below 10$^{20.3} \rm{cm}^{-2}$, denoting the edge of the gaseous disc. This radius (\rgalStop) will be used to normalize forthcoming plots. The disc in \textbf{m12f} underwent a recent merger event with an LMC-like object at redshift~0.08, resulting in the streaming accretion feature on the right side of the plot and affecting numerous metrics in our analysis. Galactic rotation is counterclockwise in all projection plots.
}
\end{figure}

This study is based on 4 simulated Milky Way-mass discy galaxies (\textbf{m12m, m12i, m12f}, and \textbf{m12b}) evolved in cosmological context (see Fig.~\ref{fig:facePlot} for face-on and view of their gaseous discs) where gas infall, large and small scale outflows, and galaxy interactions are modeled self-consistently. Relevant properties for each galaxy can be found in Table~\ref{table:discs}. Note, the extent of the gaseous disc is larger than the stellar disc in all cases.
Simulations utilize a "zoom-in" technique to reach high resolution in fully cosmological settings and were run with the gravity+(magneto)hydrodynamics code GIZMO \citep{hopkins15} using Lagrangian Godunov meshless finite mass method.

\begin{table*}
\centering
\begin{tabular}{p{1.25cm} p{1.25cm} p{1.25cm} p{1.25cm} p{1.25cm} p{1.25cm} p{1.25cm} p{1.25cm} p{1.25cm} p{1.25cm} }
\hline
Simulation & $R_{\rm vir}$ & $R_{*,1/2}$ & \rgal &  $R_{\rm HI,19}$  & $h_{\rm total}$ & $h_{\rm{cold,inner}}$ & \tgal & $v_{\rm c}$ & $M_{*}$ \\
Name & [kpc] & [kpc] & [kpc] & [kpc] & [kpc] & [kpc] & [Gyr] & [km s$^{-1}$] & [$M_{\odot}$] \\
\hline
\hline
\textbf{m12m} & 232.0 & 7.84 & 26.8  &  30.5  & 0.87 & 0.21 & 0.867 & 190 & 3.4e10\\
\textbf{m12i} & 215.4 & 3.61 & 17.1 &  20.6 & 0.76 & 0.18 & 0.591 & 178 & 2.4e10\\
\textbf{m12f} & 237.1 & 3.72 & 18.3 &  28.8 & 1.01 & 0.33 & 0.583 & 193 & 3.4e10\\ 
\textbf{m12b} & 221.2 & 1.81 & 11.7 &  15.1 & 0.47 & 0.14 & 0.349 & 206 & 3.0e10\\ 
\hline
\hline

\end{tabular}

\caption{\label{table:discs} Parameters characterizing the size of the disc for the four galaxies in our sample at z=0. $R_{\rm vir}$ is the virial radius (calculated following \citet{bryan98}). $R_{*,1/2}$ is the radius at which half the stellar mass is contained. \rgal is the radius at which the total hydrogen column density drops below $10^{20.3} \rm cm^{-2}$ when viewed face-on, signifying the transition to a column density below a Damped Lyman Alpha (DLA) system. Likewise, $R_{\rm HI,19}$ is the radius at which the HI column density drops below $10^{19} \rm cm^{-2}$. $h_{\rm total}$ is the scale height of the total gas and $h_{\rm cold,inner}$ is the scale height of the cold hydrogen ($T <$ 100 K) in the inner 5 kpc. Scale height was calculated as the height at which the average gas density drops by a factor of \textit{e} from the average value within $\pm$20 pc of the midplane. The parameter \tgal is the dynamical time of the galaxy, defined \tgal=$2\pi$\rgalStop $/v_{\rm c}$. The rotational velocity ($v_{\rm c}$) is the value predicted from the enclosed mass at 0.5 \rgalStop. Full rotation curves for most galaxies in our sample can be found in \citet{hopkins20}. $M_{*}$ is the stellar mass contained within 3 $R_{*,1/2}$.}
\end{table*}

Cooling, star formation, and stellar feedback are implemented as in FIRE-2 \citep{hopkins18}, an updated version of the original FIRE project \citep{hopkins14}. The simulations include photoionization and photoheating from the cosmic UV background based on the \cite{faucher09} model.
Star formation is enabled in self-shielding/molecular, Jean unstable, dense ($\rm n_{\rm H} > 1000 \rm cm^{-3}$) gas. 
Once created, star particles are treated as single-age stellar populations with IMF-averaged feedback properties calculated from STARBURST99 \citep{leitherer99} assuming a \cite{kroupa01} IMF. Feedback from SNe (Type Ia and II), stellar mass loss (O/B and AGB), and radiation (photo-ionization, photo-electric heating, and UV/optical/IR radiation pressure) are treated as in \citet{hopkins18}. 
In this analysis, we use simulations that in addition to standard FIRE-2 feedback incorporate magneto-hydrodynamics (MHD) and cosmic ray (CR) physics from \cite{hopkins20} using CR transport methodology described in \cite{chan19}. To summarize, runs include CR injection in SNe shocks, fully-anisotropic CR transport with streaming, advection, and diffusion, CR losses (hadronic and Coulomb, adiabatic, streaming), and CR-gas coupling. These runs adopt a constant CR diffusion coefficient. Physically motivated transport models that allow variations of the effective diffusion coefficient can give a variety of behaviors in between these constant diffusion coefficient runs and runs without CRs \citep{hopkins_21_crTransport}.

We provide a brief overview of the histories, morphology, and dynamics of these four galaxies. For more details, please see the provided references. The system \textbf{m12m} is an earlier forming halo with a large, more broadly distributed stellar disc; \textbf{m12i} is a later forming halo that forms a massive disc, with a minor merger near z$\sim$0.7 and two fly-bys at z$\sim$0.3-0.2; \textbf{m12f} is in the late stages of a merger event (z$\sim$0.1) with an LMC-mass object, which has changed the dynamics of the outer regions of this galaxy both in terms of radial velocities and torques; \textbf{m12b} is an earlier forming halo with a compact bulge and thin disc \citep{hopkins20}.

Broadly speaking, while some FIRE-2 simulations form bar structures within the inner regions of Milky-Way mass galaxies, they tend to be weaker and shorter-lived than what is seen in other simulations \citep{sioree23-Fire2-Bars}. 
The only galaxy considered in this study that experiences any bar formation in its history is \textbf{m12b}.

As shown in \cite{trapp22}, the most significantly consistent source of fueling to star-forming regions at late times for these discs are radial flows largely parallel to the disc plane. In the iCGM, these radial speeds can be up to $\pm$100 km/s with a time-averaged inflow rate of 10-20 km/s. Within the disc, the radial speeds can still show deviations up to $\pm$40-60 km/s but averaged inflow rates are around 1-5 km/s. In both regions, the averaged radial mass flux rate tends to be around 2-3 \MFunit. This is not to say, however, that the vertical flows are not present or insignificant in these systems. The CR+ galaxies show strongly collimated central outflows with velocities up to 100-150 km/s \citep{hopkins21_crOutflows}. Significant vertical velocities (40-60 km/s) can be seen throughout the discs of these systems, however, the net mass flux rate of these vertical flows is not as consistently significant as the parallel flow rates \citep{trapp22}. In some systems at some times (e.g. particularly \textbf{m12i} at redshift z$\sim$0.1) the vertical mass flux rate can be of a similar order to the parallel mass flux rate. During the redshifts considered in this study, net vertical mass fluxes tend to be an order of magnitude less than radial inflows, with \textbf{m12f} and \textbf{m12b} showing net vertical inflow and \textbf{m12m} and \textbf{m12i} going through episodes of net vertical inflow and outflow. Also note, that while gas trajectories in the iCGM are largely parallel to the disc on average, gas tends to drop vertically downward just prior to joining in the outer half of the disc. For a more detailed analysis of vertical outflows in the FIRE simulations, see \cite{hopkins21_crOutflows,chan22} for the runs with the additional CR physics considered here, and \cite{pandya21,porter24} for runs without additional CR physics.

 The FIRE simulations have been successful in matching a range of galactic properties to observations, including $M_{*}/M_{\rm{Halo}}$; star formation rates and histories \citep{hopkins14,sparre17,feldmann16,santistevan20}; dense HI covering fraction in the circumgalactic medium at both low and high redshift \citep{faucher15,faucher16,hafen17}; outflow properties \citep{muratov15, muratov17,alcazar17,pandya21}; metallicities \citep{ma16a,bellardini21,bassini24,marzewski24}; morphological/kinematic structure of thin/thick discs \citep{ma17b,yu21,sanderson20}; baryonic and dark matter mass profiles and content within the halo \citep{chan15, wetzel16}; Giant molecular cloud properties \citep{oklopcic17:clumps_in_fire,benincasa20,guszejnov20}; and circular velocity profiles \citep{hopkins18}.

 In general, our MW-mass galaxies with CRs show good agreement with the observationally inferred M$^{*}$-M$_{\rm{halo}}$ relation and are disc dominated \citep{hopkins20}. Stellar masses in runs without CRs are somewhat higher than this relation predicts. On the other hand, CR+ runs may be slightly underestimating stellar mass, although statistics are poor and both are within observational uncertainties.
 Our choice of simulations with additional CR physics is mainly guided by their lower late-time star formation rates of 2-3 $\rm M_{\odot}/yr$ vs 3-10 $\rm M_{\odot}/yr$ in FIRE-2 simulations without CRs, which is closer to what is seen in the Milky Way.  This lower star formation is associated with lower velocity dispersion of galactic discs \citep{chan22} and potentially additional planar alignment of the accreting gas \citep{hopkins21_crOutflows}. In addition, the circum-galactic medium in MW-mass simulations with CRs agrees with observations of low and intermediate ions seen in absorption systems around galaxies \citep{ji20}. We previously found that both default FIRE-2 simulations and CR+ runs show similar {\it qualitative} behavior and trends in gas accretion onto low redshift discs. The inclusion of additional CR physics in these simulations provides significant non-thermal pressure support that confines infall geometry closer to the galactic plane and slows infall rates, but the overall picture of gas co-rotation as it approaches the disc remains the same \citep{hopkins21_crOutflows, trapp22}.

The starting baryonic element mass in our simulations is $\rm m_{b,min} = 7100 M_{\odot}$ and a typical gravitational force softening for star-forming gas is $\sim 2$ pc. Note that spatial resolution (softening and smoothing lengths) for our gas elements is adaptive; typical force softening for ISM gas is $\sim 20$ pc. All simulations employ a standard flat $\Lambda$CDM cosmology with $\rm h \approx 0.7$, $\rm \Omega_{M} = 1-\Omega_{\Lambda} \approx 0.27$, and $\rm \Omega_{b} \approx 0.046$ (consistent with \cite{planck14}).

\begin{figure}
    \center{\includegraphics[width=.45\textwidth]
 	       {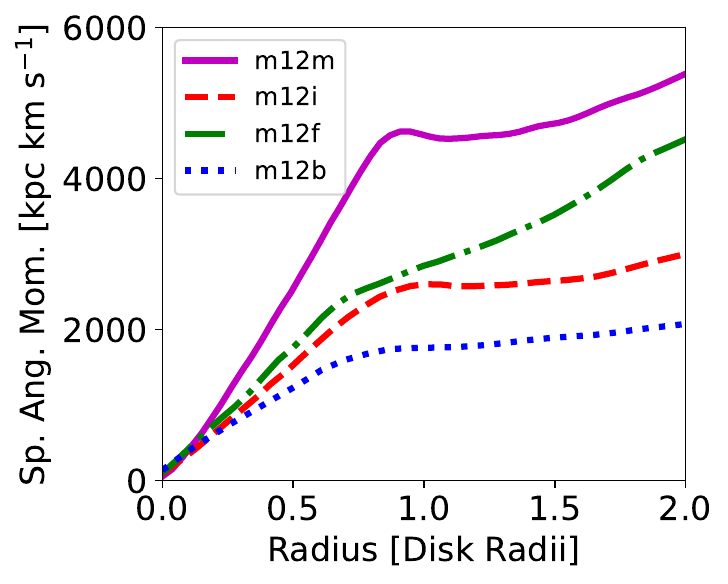}}
	  \caption{\label{fig:jPlot} Mass-weighted average specific angular momentum of all gas within $\pm$10 kpc of the disc plane as a function of cylindrical radius, normalized by the gaseous disc radius (\rgalStop, Table~\ref{table:discs}). Values were averaged over the 10 snapshots ($\sim$200 Myr) considered in this study. Angular momentum increases with radius within the disc and decreases slightly at the disc edge. The slope of the angular momentum curve is shallower in the nearby CGM, but still positive. As shown in \citet{trapp22}, the angular momentum of the gas within the disc is strongly aligned in the +z direction.
}
\end{figure}

We focus on the gas at late times, when our simulated galaxies have discs that are thin, stable, and have a clear orientation. The specific angular momentum of the gas in each galaxy can be seen in Fig.~\ref{fig:jPlot}. To isolate the various sources of torque, we restart each simulation from a snapshot at redshift z$\sim$0.02. We modified the output for the FIRE-2 CR+ code to track and output the gravitational accelerations, the total hydro accelerations, a breakdown of hydro contributions from the Riemann solver (pressure\footnote{The pressure gradients analyzed include the Cosmic Ray pressure \citep{chan19} and magnetic pressure terms, however we consider them separately as well.}, magnetic tension, and non-continuum terms), viscous forces, and accelerations from radiative transfer. We additionally track and output the net angular momentum change between snapshots arising from these forces, as well as from the direct injection of momentum and angular momentum from supernovae and stellar winds. We save 10 snapshots with a time-stepping of $\sim$20 Myr as was done in the original simulation run.

For each galaxy, we define a Cartesian coordinate system centred on the galactic centre with the z direction oriented along the angular momentum vector of the galaxy. We determine the orientation of the angular momentum vector from the vector sum of angular momenta of cold dense gas (T $<$ 8000 K, $n > 1 \rm cm^{-3}$ ) in the inner 10 kpc, with respect to the galactic centre. We calculate the galactic centre from the mass distribution of the star particles using a shrinking sphere algorithm.\footnote{A 1 Mpc radius sphere was defined around the region of maximal gas density and the stellar centre of mass was calculated. This sphere was shrunk by a factor of 0.7 and the centre of mass was recalculated until it reached a radius of 10 kpc. The centring was repeated without shrinking until the centre converged to a stable value.} We determine the galactic velocity by calculating the mass-weighted velocity average of all-star and dark matter particles within 15 kpc of the galactic centre.

\subsection{Calculation of Torques}

We characterize the torques acting on gas using two methods: outputting the acceleration values from each force at every snapshot and outputting the corresponding net change in angular momentum from the previous snapshot due to said force. In the first method, used primarily for visualizations at a single time point, torques are calculated as follows:

\begin{equation} \label{eq:1}
     \boldsymbol{\tau}_{j} = \frac{1}{M_{\rm{bin}}} \sum_{i} \big{[}m_{i} \boldsymbol{r}_{i} \times \boldsymbol{a}_{j}(\boldsymbol{r}_{i})\big{]}
\end{equation}

wherein $M_{\rm{bin}} = \sum_{i}m_{i}$ is the sum of the masses in the spatial bin (see Sec.~\ref{sec:methodology-bins} for bin definition), $\bf{a}_{j}(\bf{r}_{i})$ is the acceleration due to the force $j$ acting on particle $i$.
Calculations were done in the galactic frame as described in the previous section. This acceleration is taken directly from the acceleration values at a given snapshot time as used in the simulation code.
Note, in this formulation of torque there would also exist a Lagrangian torque $\dot{m_{i}} \boldsymbol{\ell}$ that refers to the change in the numerical element mass due to enrichment from stellar feedback. Given that these simulations are run in MFM (meshless finite mass) mode, there is no additional mass transfer between hydrodynamical cells. We have confirmed that this term has little to no effect on the specific torques considered throughout this study and is not displayed.

The second method involves directly calculating the torque during the simulation run and summing the resulting changes to gas velocity to determine the change in specific angular momentum between snapshots. At each timestep in the simulation where an element is updated, the change in specific angular momentum for a given force is calculated as follows:

\begin{equation} \label{eq:2}
    \Delta \boldsymbol{j}_{i} = \sum_{t} (\boldsymbol{r}_{i,t} - \boldsymbol{r}_{C,t}) \times \Delta \boldsymbol{v}_{i,t}       
\end{equation}

wherein $(\boldsymbol{r}_{i,t} - \boldsymbol{r}_{C,t})$ is the position vector of the element relative to the galactic centre at time $t$ and $\boldsymbol{v}_{i,t}$ is the velocity kick the given force is imparting on the element at that timestep. Given that calculating a galactic centre on the fly is computationally expensive for the number of updates being done, during the reruns we determine the centre value at a given simulation timestep by linearly interpolating the previously calculated centres from the initial run at each snapshot output (see last paragraph of Section~\ref{sec:methodology-galaxySample}). After writing out a snapshot, this $\Delta \boldsymbol{j}_{i}$ is reset to 0 for each gas mass element. The average torques are then calculated by dividing by the time-step between snapshots.

This technique has two primary advantages. Firstly, it allows us to fully recreate how the angular momentum is changing between snapshots, instead of relying on acceleration vectors at a single time point, which may be fluctuating. Secondly, this allows for a more direct comparison between all sources. Torques from stellar feedback can only be calculated in this way, as they do not have a corresponding acceleration variable. A direct comparison between the total change in specific angular momentum from the sum of these torques and the change as measured by the difference between two snapshots is shown in Fig.~\ref{fig:djdtScatter} and follows a tight linear relationship.

The subsequent results and analysis focus on the component of the torque along the direction of the angular momentum vector of the galaxy (see Sec.~\ref{sec:methodology-galaxySample} for definition). In the following subsections, we will discuss in more detail each different source of torque considered.

\subsubsection{Gravitational Torques}

We consider the torque from the net gravitational acceleration on each gas mass element, as well as the contributions from gas, stars, and dark matter separately. We calculate the net gravitational torques both from the acceleration vectors at a given snapshot (Eq.~\ref{eq:1}) and from the net change in specific angular momentum (Eq.~\ref{eq:2}) divided by time between consecutive snapshots.
Individual contributions from gas, stars, and dark matter, were calculated separately in post-processing as follows:
\begin{equation}
\bf{a}_{j} = \nabla \Phi_{j}
\end{equation}

where $\Phi$ is the gravitational potential from solely the particle type of interest and was estimated using M. Grudic's \textsc{pytreegrav}\footnote{https://github.com/mikegrudic/pytreegrav} with a tree opening of $\theta=0.7$.

\subsubsection{Magnetohydrodynamical Torques \label{sec:methodology_MHD_Torques}}

We directly extract the MHD torque from the HLLD Riemann solver, separating the usual continuum/adiabatic contributions from thermal pressure, CR "pressure", magnetic pressure, and magnetic tension, as well as the "non-continuum" terms. These non-continuum terms include the Reynolds stress, anisotropic corrections to the CR pressure, and numerical/upwinding (analogous to a bulk viscosity when two gas cells try to cross) terms that arise from the Riemann problem (see \cite{hopkins16_AccurateMeshlessMethodsForMHD} Sec. 2.1 and 2.2 for more details). We extract the MHD torques from the Riemann problem directly as opposed to a post-processing estimate of the pressure gradient, as this cannot include the exact numerical definition of the gradient, shocks, nor numerical dissipation terms.

Viscous fluxes were calculated via an implementation of anisotropic Spitzer-Braginskii viscosity, which adds an anisotropic viscous stress-energy tensor to the momentum and energy flux (see \cite{hopkins18}, Eq. F3). These contributions to the MHD fluxes are then divided by the element mass to get the corresponding accelerations. As with the gravitational torques, all MHD torques are then calculated as described in both Eq.~\ref{eq:1} and Eq.~\ref{eq:2}.

\subsubsection{Torques from Radiative Transfer}

The FIRE-2 simulations include an implementation of radiative feedback for radiation pressure, photoionization, and photoelectric effects through the radiative transport algorithm "LEBRON" (Locally Extincted Background Radiation in Optically-thin Networks). Each star particle is treated as a source of radiation with appropriate age and metallicity-dependent, IMF-averaged spectrum (\cite{hopkins18}, Appx. E). We calculate torques from this radiative transfer as in both Eq.~\ref{eq:1} and Eq.~\ref{eq:2}. \footnote{The accelerations from these radiative transport terms reduce to $\textit{\textbf{a}}_{\rm{rad}}=\kappa_{\nu} \textit{\textbf{F}}_{\rm{\nu}}/c$. in the optically thin limit, where $\kappa_{\nu}$ is the flux weighted opacity in each band and $\textit{\textbf{F}}_{\rm{\nu}}$ is the incident flux. See \cite{hopkins18}, Appx. E.}

\subsubsection{Torques From Stellar Feedback}
The FIRE-2 simulations additionally incorporate mechanical feedback from SNe (Types Ia \& II) and stellar winds. Important in our consideration of changes in angular momentum is the injection of momentum into surrounding gas mass elements. Unlike the other sources of torque considered, this term is not treated as an acceleration but is rather directly added to the velocity term of the gas. It will therefore not be accounted for in the MHD accelerations.

To briefly summarize, at every timestep for each star particle, it is determined whether an event occurs (SN Ia, SN II, or non-zero stellar mass loss). If an event does occur, the surrounding gas elements are identified and their "effective faces" that would be seen by the star particle are determined. The ejecta is then integrated over a solid angle through to each face and the momentum flux is assigned in the rest-frame of the star. This injected momentum is then boosted back to the simulation frame. Before being added to the gas mass element, the code accounts for the $PdV$ work. 
Importantly, this implementation of momentum injection maintains the conservation of linear momentum (as well as mass and energy), with the ejecta in the rest frame of the star particle unbiased in one direction or another even in regions with anisotropic gas distributions. This direct momentum injection can be converted to a change in angular momentum as described in Eq.~\ref{eq:2}. Because mass elements receiving momentum are at different galactic radii, certain elements may experience larger torques for a given amount of linear momentum injected.

Note that the torques from SNe or other forms of stellar feedback that we define and discuss throughout are an explicitly {\em resolution-dependent} concept. If we had infinite resolution, this “feedback torque” would be identically zero, as it would come strictly from the initial ejecta (which we assume emerges isotropically from a point source, and such has zero net angular momentum).  However, the vast majority of the radial acceleration/momentum (and all of the torque/angular momentum) imparted over the course of the expansion of a SNe blastwave, superbubble, or stellar wind bubble comes from the $P dV$ work imparted on the {\em ambient} gas during the Sedov-Taylor and early snowplow phases of expansion. Since this region has a finite extent ($\sim$1-10 pc) and an inhomogeneous density distribution, some angular momentum change will necessarily be imparted to those gas elements. This would therefore, at infinite resolution, appear entirely in the hydrodynamic torques we define (specifically in the “pressure” torque term). We are restricted to finite resolution, and for Milky-Way mass galaxies in fully cosmological settings, it is still generally impossible to well-resolve the cooling/Sedov-Taylor radii of individual SN remnants let alone the initial pure ejecta (which requires sub-solar mass resolution). As such, the prescription in FIRE-2 couples ejecta momentum from SNe and stellar mass-loss to the resolved gas elements/cells in the simulation at their centre-of-mass locations, integrating over a standard analytic Sedov-Taylor solution (calibrated explicitly to orders-of-magnitude higher resolution simulations of individual SNe remnant expansion) to estimate the un-resolved $PdV$ work \citep[see][for details]{hopkins18b-SneMethods}. Thus we account for this explicitly in its own torque “budget”. This finite-resolution effect gives an unexpected advantage, however. Because the cooling radii of individual remnants are largely unresolved, this allows us to approximately separate the contribution to MHD torques driven more ``directly’’ by stellar feedback, from that arising entirely from non-feedback sources (e.g.\ gravito-turbulent stresses).

\subsection{Analysis Techniques}

\subsubsection{Spatial Averages} \label{sec:methodology-bins}
To visualize the torques acting within a galaxy, we spatially average the z-component of the torques on each gas mass element. For creating projection maps, we first calculate the mass-weighted averages of the specific torques as determined from the acceleration vectors at a given snapshot (see Eq.~\ref{eq:1}) in 0.5x0.5 kpc$^{2}$ cartesian bins within $\pm$10 kpc of the disc plane. The choice of this height cutoff is motivated by the fact that the mass-weighted average torques in the disc are not very sensitive to this cutoff, but it allows us to capture the accreting gas in the iCGM, which can have a more flared structure. Note, for visualizations of the torques arising from SNe and stellar winds, Eq.~\ref{eq:2} is the only option, as there is no corresponding acceleration vector. The in-plane component of the torque vector (e.g. x- and y-), which on a disc averaged sense will correspond to the alignment/misalignment of gas with the plane, is beyond the scope of this study and is not considered here. 

We additionally calculate the mass-weighted averages as a function of cylindrical radius. For these calculations, the individual timestep changes in specific angular momentum summed over the interval between snapshots and divided by the snapshot time interval (Eq.~\ref{eq:2}) were used in all cases to make sure no quantitative information was lost due to temporal resolution and to allow for more direct quantitative comparisons between all sources of torque.

\subsubsection{Particle Tracking}
For a more detailed analysis of how torques evolve for individual gas mass elements, we look at the torque evolution of individual mass elements across the 10 snapshots considered. We define the edge of each disc as in \cite{trapp22}, wherein \rgal is the radius where the average hydrogen column density drops below 10$^{20.3} \rm{cm}^{-2}$. We focus our analysis on three distinct populations of mass elements: elements with an average distance within r $\leq$ $\frac{1}{2}$ \rgal from the galactic centre (Inner Disc), elements with an average distance between $\frac{1}{2}$\rgal < r $\leq$ \rgal (Outer Disc), and elements with an average distance between \rgal < r $\leq$ 4 \rgal (inner CGM, iCGM). For each gas mass element, the various torque sources as determined in Eq.~\ref{eq:2} are summed across the 10 snapshots to determine its net contribution.

\section{Results} \label{sec:results}

\subsection{Torques as a function of cylindrical radius}

We start by analyzing the torques on the gas mass elements as a function of cylindrical radius arising from the three dominant sources considered: gravity, hydro forces, and momentum injection from SNe. All radial plots show mass-weighted averages within $\pm$10 kpc of the disc plane.

\begin{figure}
	  \center{\includegraphics[width=.5 \textwidth]
 	       {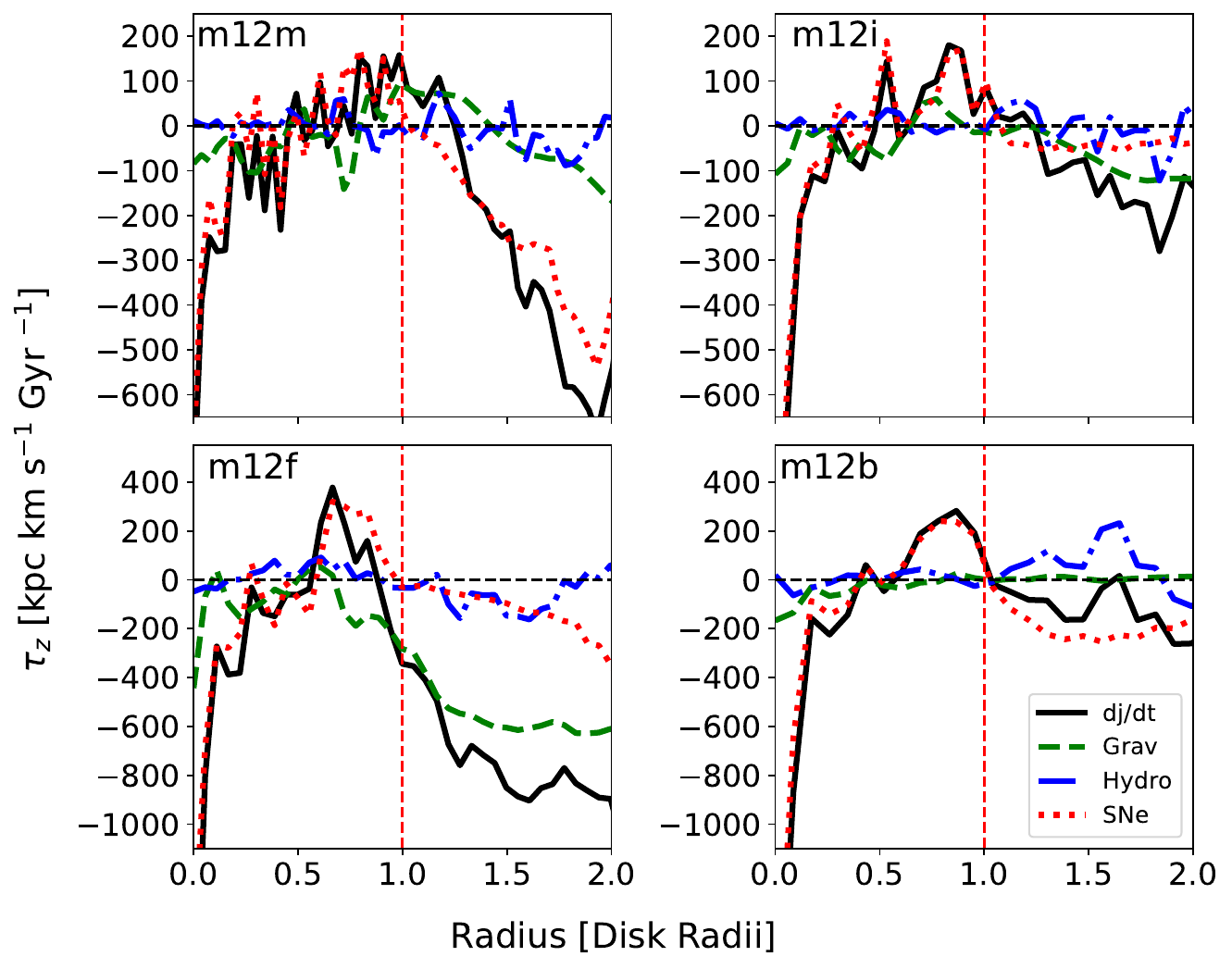}}
	  \caption{\label{fig:totalTorqueLinePlot} Specific torque acting on gas as measured from summed changes in angular momentum during the run (black, solid line) as a function of cylindrical radius as well as the three main sources: gravitational torques (green, dashed line), hydrodynamical torques (blue, dash-dot line), and feedback torques from SNe (red, dotted line). Each panel shows a different galaxy. Plots are averaged over 10 snapshots ($\sim$200 Myr) near redshift z=0. The radius has been normalized to the size of the gaseous disc (\rgalStop). Values are averaged between $\pm$10 kpc from the disc plane. The horizontal black dashed line marks 0 torque. Negative values correspond to torques that drive inflow. Both the gravitational torques and hydrodynamical torques tend to be negative within the inner disc and more positive in the outer disc, with the gravitational torques being slightly more significant on average. Gravitational torques are negative in the inner CGM (iCGM), due to dark matter structure. The residual torques from SNe show strong negative values in the most interior regions, where the majority of star formation is happening, positive values further out in the disc, and weaker negative values in the iCGM. For gas within these discs moving radially inward at velocities of $\sim$1-5 km s$^{-1}$, the corresponding angular momentum transfer for these systems would be $\sim$150-750 kpc km s$^{-1}$ Gyr$^{-1}$.
}
\end{figure}

In Fig.~\ref{fig:totalTorqueLinePlot} we display the mass-weighted average specific torque as calculated by the change in specific angular momentum during the simulation run (Eq.~\ref{eq:2}). We also show the mass-weighted average specific torques from total gravitational forces, total hydro forces, and SNe. The gravitational torques tend to be negative in the inner disc and tend to be more positive in the outer disc. In the iCGM, the gravitational torques turn over to a smoother inward torque. In \textbf{m12f}, the larger gravitational torques in the iCGM are related to the merger event. Even though torques are as significant, overall angular momentum is much higher in the iCGM (Fig.~\ref{fig:jPlot}), so this is not causing dramatic changes. The hydro torques switch between positive and negative torques within the disc. Although difficult to see in this plot, they contribute a net negative torque within the disc in most cases, and a net positive torque just outside of the disc edge and within the inner few kpc of the disc. The torques arising from the sub-grid $PdV$ work done by SNe remnants in their expansion are significant throughout the disc and iCGM. They provide strong negative torques in the innermost regions of the disc, positive torques in the outer regions of the disc, and negative torques throughout the iCGM. Note, \textbf{m12m} has the most extended star formation, and thus these SNe torques are more significant in the iCGM compared to other galaxies in our sample.

\begin{figure}
	  \center{\includegraphics[width=.5 \textwidth]
 	       {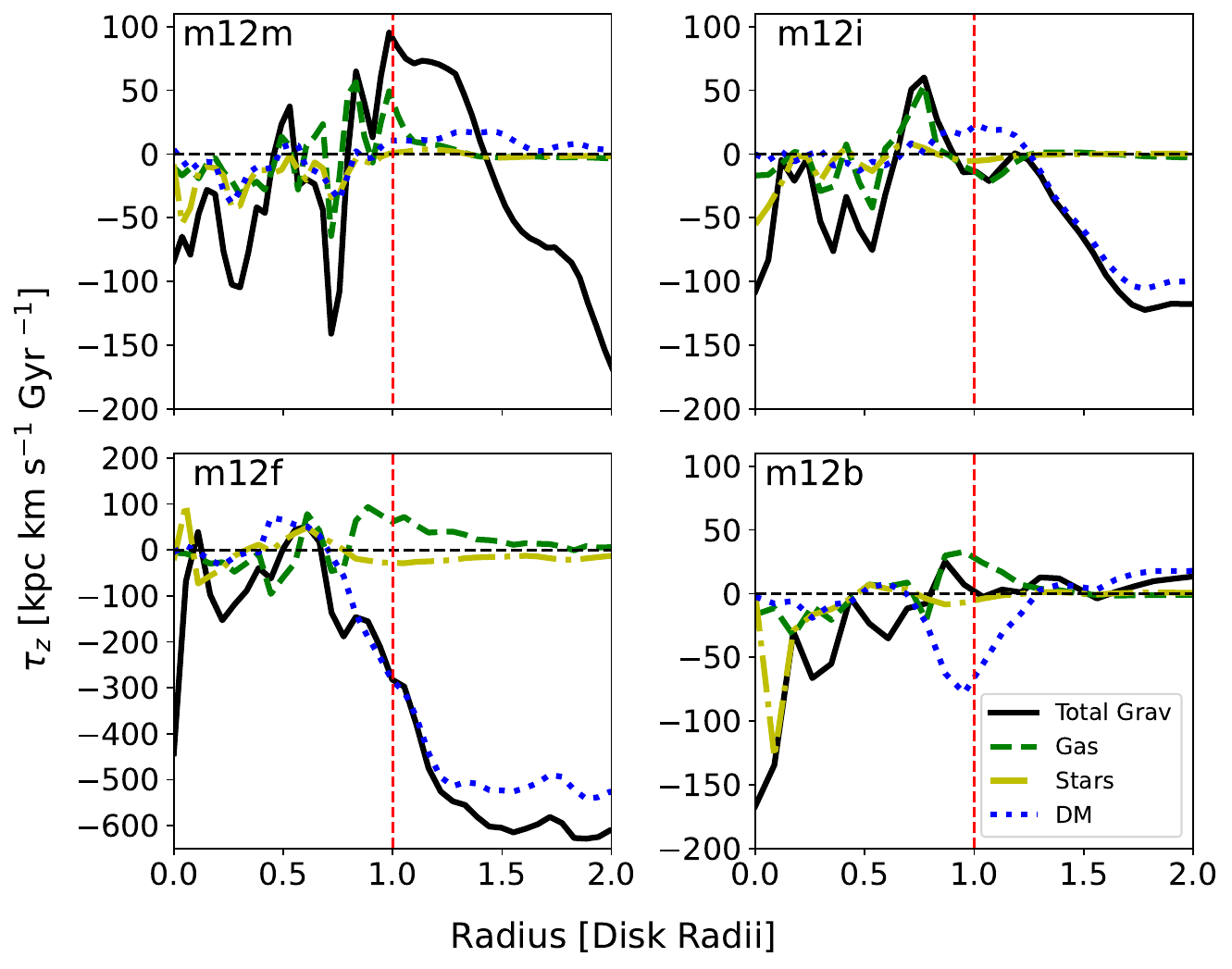}}
	  \caption{\label{fig:gravTorqueLinePlot} Gravitational specific torques acting on gas averaged over 10 snapshots ($\sim$ 200 Myr) as a function of cylindrical radius from the three particle types in the simulation: gas (green dashed), stars (yellow dash-dot), and dark matter (blue dotted). Subcomponent contributions were estimated in post-processing. Therefore, values do not match up perfectly but can be viewed as an estimate of which source is dominant at that radius. Torques from the stars tend to be dominant in the inner disc ($\lesssim$5 kpc) and drive a net angular momentum loss in most cases. Gas-gas gravitational torques tend to be more dominant in the outer regions of the disc. There tends to be a larger positive torque feature just interior to the disc edge, roughly corresponding to the point where gas mass elements are predominately joining the disc. The torques from the dark matter are not insignificant within the disc but tend to become more relevant at larger radii, providing a negative torque in the iCGM in most cases. Note the different scale for \textbf{m12f} in order to capture stronger torques owing to the recent merger event.
     }
\end{figure}

We further break the gravitational forces down in Fig.~\ref{fig:gravTorqueLinePlot}, showing the contributions of the gravitational effects from the gas, stars, and dark matter acting on the gas mass elements.\footnote{Note that the relative contributions were calculated at a single time point in post-processing, while the total gravitational torques were calculated from the summed net angular momentum change between snapshots ($\Delta$t=20 Myr), so they do not match precisely.} Torques from the stars tend to be the most significant gravitational torques within the innermost regions of the disc, while torques from the gas tend to be more significant within the outer disc. Torques from the dark matter, while not insignificant within the disc, become relatively more significant near the disc edge and into the iCGM, where gaseous and stellar torques drop off. In all cases except \textbf{m12b}, there is a turnover in the iCGM where the gravitational torques become a source of negative torques due to the triaxial dark matter halo structure \citep{shen21-DmHaloesInFIRE}. The dark matter torques arise primarily from the quadrupole component of the potential \citep{bowden13-TriaxialHaloes}, which we discuss in Sec.~\ref{sec:results-visualizing-torque}. The relative contributions from gas, stars, and dark matter largely stem from their distribution in these galaxies. In all cases, the distribution of gas drops off sharply at the disc edge \citep{trapp22}, and the scales of the stellar discs are even smaller (Table~\ref{table:discs}). A few notable differences between these galaxies are the extended stellar disc of \textbf{m12m}, which leads to more significant torques from stellar gravity at larger radii, the merger event in \textbf{m12f} which leads to much stronger dark matter torques in the iCGM, and the more compact nature (and history of weak bars) in \textbf{m12b}, which influences the strength of the stellar torques in the innermost regions.

\begin{figure}
	  \center{\includegraphics[width=.5 \textwidth]
 	       {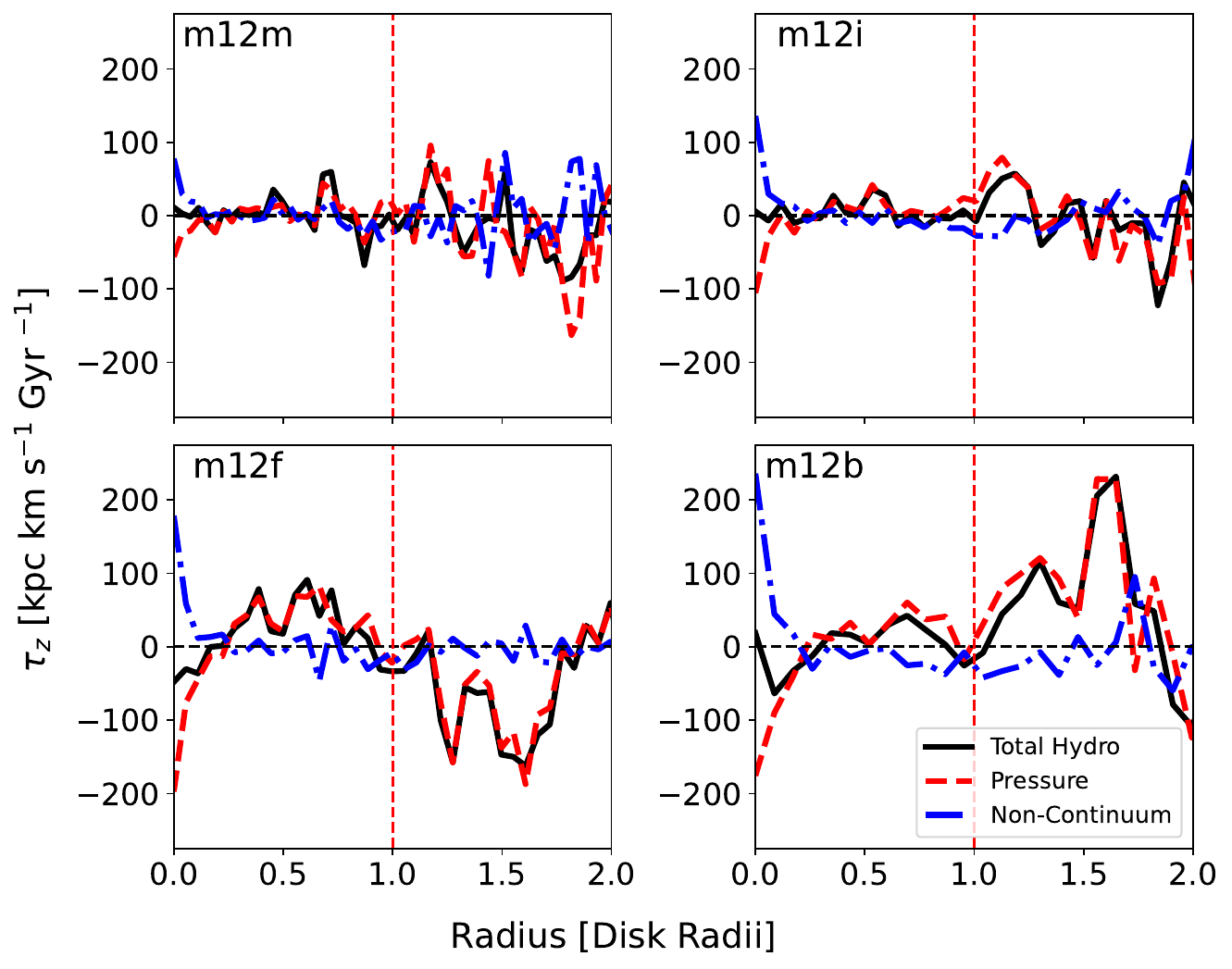}}
	  \caption{\label{fig:hydroTorqueLinePlot} Hydrodynamical specific torques acting on gas averaged over 10 snapshots at z=0 ($\sim$ 200 Myr) as a function of cylindrical radius. The red dashed line shows the specific torque from the pressure gradients (thermal+CR) on the gas, read out directly from the Riemann Solver within the simulation code. The blue dash-dotted line shows the contribution from the non-continuum terms in the Riemann solver (see Sec.~\ref{sec:methodology_MHD_Torques}). The pressure and non-continuum terms tend to be of a similar order of magnitude within the disc, with the pressure torques slightly dominant. The pressure torques show negative values in the innermost regions of the disc, and positive values in the outer half of the disc, similar to the SNe torques. The pressure torques additionally peak just outside the disc edge in most cases. The non-continuum terms typically work opposite the pressure terms, limiting the overall effect of the hydrodynamical torques.
}
\end{figure}

In Fig.~\ref{fig:hydroTorqueLinePlot}, we further separate the hydrodynamical torques into the usual continuum/adiabatic terms in the "pressure" torque, which includes CR "pressure" forces and thermal pressure with their continuum gradients, and the remaining terms in the non-continuum torque (see Sec.~\ref{sec:methodology_MHD_Torques}). Torques from magnetic pressure, magnetic tension, and viscous shearing are largely subdominant and, for simplicity, considered separately (see Appx.~\ref{sec:appendix_misc}).
The individual terms tend to vary as a function of radius more strongly than the gravitational torques, averaging out slightly when summed together. The pressure and non-continuum torques are of a similar order of magnitude within the disc, although the pressure term tends to be more significant. Additionally, the pressure torques tend to be the dominant hydrodynamical term in the CGM. Of particular note is the tendency for the pressure torques to show a peak in positive torques just outside the disc edge, most saliently in \textbf{m12i} and \textbf{m12b}. We further break down the contributions of the pressure torques into thermal, CR, and magnetic pressure in Fig.~\ref{fig:pressureSubcomponentRadialPlots}.

\subsection{Visualizing Torques Structure}\label{sec:results-visualizing-torque}
\begin{figure}
	  \center{\includegraphics[width=.5 \textwidth]
 	       {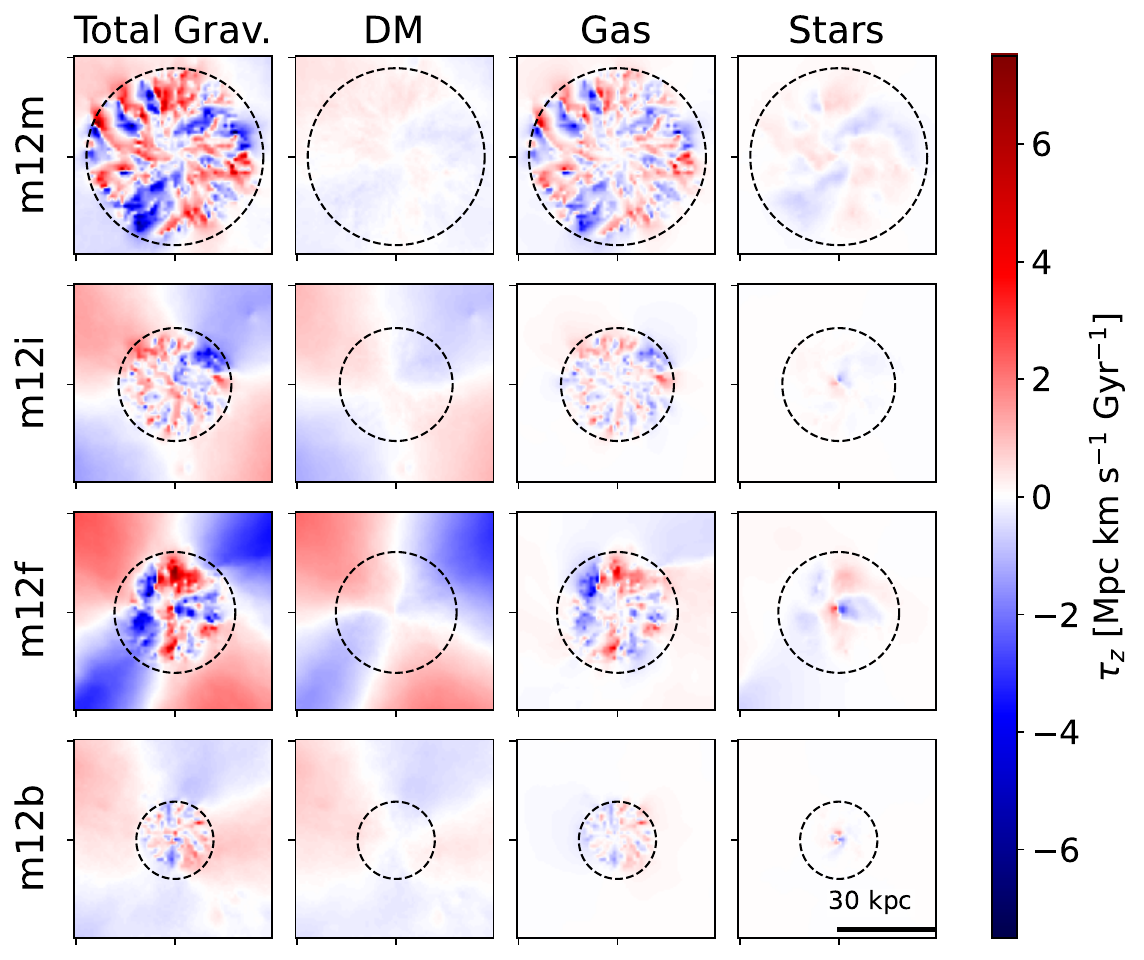}}
	  \caption{\label{fig:gravTorqueFacePlot} Face-on projections showing the mass-weighted average specific torques from gravity acting on gas mass elements at redshift z=0 for the four galaxies of interest. Positive torques (red) roughly correspond to outflow, while negative torques (blue) roughly correspond to inflow. The dashed circles show the edge of the gaseous disc. The leftmost column shows the total gravitational torques, calculated from the simulation's gravitational acceleration vector. The right three columns show the gravitational torques caused by the dark matter, gas, and stars on the gas elements estimated in post-processing. The dark matter torques resemble an $\ell$=2 spherical harmonic in all cases and are the dominant contribution in the outskirts. The mode of the stellar torque fields reflects the dark matter structure, although they are much more concentrated in the centre and not as clean. The torques from the gas are of a higher mode.
}
\end{figure}

\begin{figure}
	  \center{\includegraphics[width=.5 \textwidth]
 	       {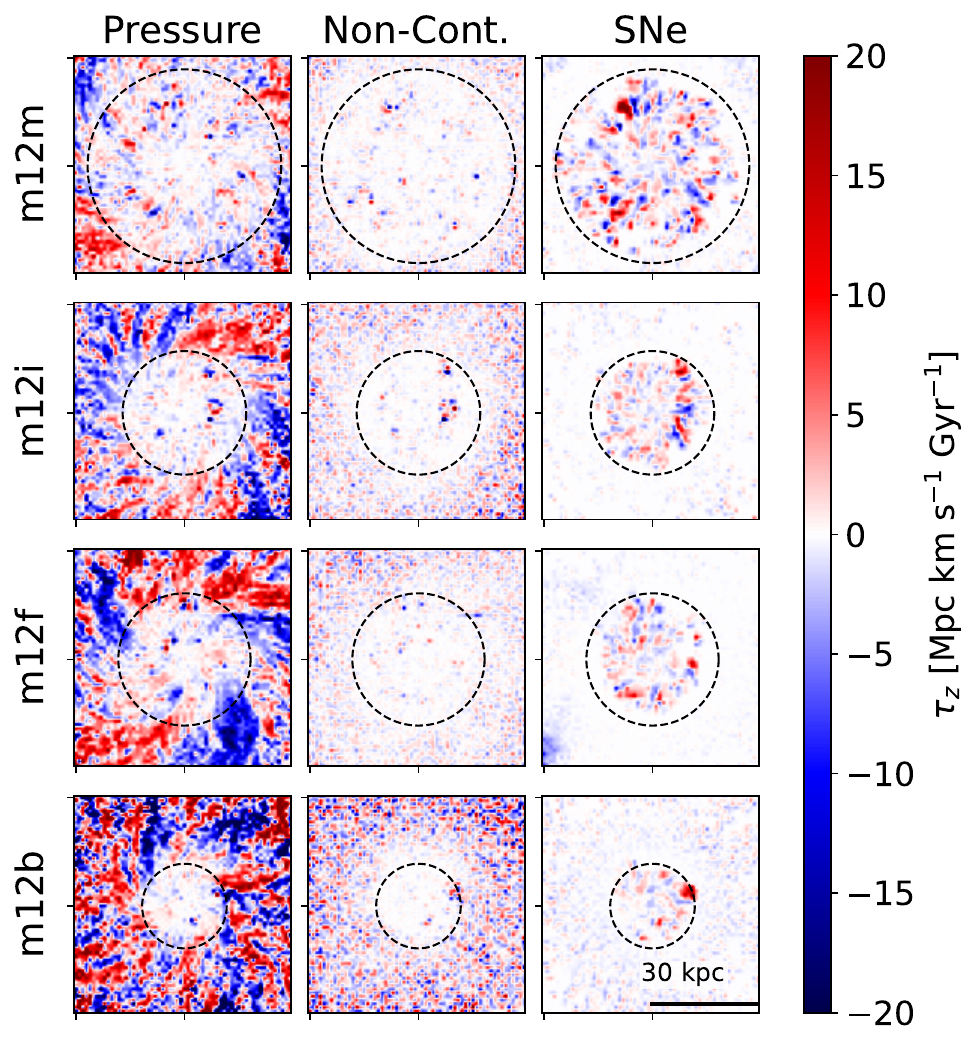}}
	  \caption{\label{fig:hydroTorqueFacePlot} Face-on projection maps showing the mass-weighted average specific torques from pressure gradients (left), the non-continuum terms in the Riemann problem (middle), and supernovae (right) acting on gas at redshift z=0. Positive values (red) correspond to outward torques, while negative values (blue) correspond to inward torques. The dashed circles show the edge of the gaseous disc. All terms are similarly significant within the disc, while the pressure term tends to be the most dominant within the CGM. 
      Gas rotationally downstream of a SNe (counter-clockwise in this orientation) is torqued outwards, and gas rotationally upstream is torqued inwards. This is expected, but the source of residual non-zero cancellation will be discussed later.
}
\end{figure}

In Figures~\ref{fig:gravTorqueFacePlot} and \ref{fig:hydroTorqueFacePlot} we present a series of face-on projections showing how the previously discussed torques are spatially structured. Positive (red) torques are causing an increase in angular momentum, roughly corresponding to outward motion. Negative (blue) torques are causing a loss in angular momentum, roughly corresponding to inward motion. Note, while previous quantitative plots show the torques as calculated in Eq.~\ref{eq:2} and averaged over multiple snapshots, these plots show only the instantaneous torques as calculated from the acceleration vectors of the simulations as in Eq.~\ref{eq:1} for one snapshot. The exceptions for this are the feedback torques from SNe and stellar winds, which can only be calculated using Eq.~\ref{eq:2}.

\subsubsection{Gravitational Torques}
In Fig.~\ref{fig:gravTorqueFacePlot} we show the structure of the gravitational torques as well as the separate contributions from dark matter, gas, and stars within $\pm$10 kpc of the disc plane.

The torque from the dark matter shows a strong $\ell$=2 spherical harmonic structure, with two distinct quadrants of negative torque and two distinct quadrants of positive torque, as would be expected from a triaxial dark matter halo \citep{bowden13-TriaxialHaloes,shen21-DmHaloesInFIRE}. These patterns evolve with time, however, they do not co-rotate with the galaxy. This implies accreting gas may pass through multiple quadrants as it joins and moves through the disc, although the azimuthally averaged torque is still negative as seen in Fig.~\ref{fig:gravTorqueLinePlot}. These torques dominate the gravitational torque in the iCGM and can be significant within the disc as well. Note that the structure and dynamics of the disc at these late times have likely influenced the anisotropies in the dark matter. This can be seen more clearly in the plot of \textbf{m12m}, which shows similar structure in the centre for the DM and stellar torque field. 

The torques from the gas-gas gravitational interactions show higher order structure, with regions of positive and negative torques of a similar spatial scale to the radial inflow and outflow structures seen in velocity in our previous study \citep{trapp22}. The significance of these torques is largely limited to the disc, dropping off sharply both outside the disc and at the very centre. Note, the sharp negative torque feature on the left side of \textbf{m12f} is due to the recent merger event.

The torque from the stars is not as significant as the other gravitational torques, implying strong symmetry of the stellar disc, but is important in the innermost regions. The structure of the torques shows a similar $\ell$=2 harmonic structure as the dark matter, with the positive/negative torquing quadrants roughly lining up in all cases. 

\subsubsection{Supernovae and Hydrodynamical Torques}
Fig.~\ref{fig:hydroTorqueFacePlot} shows the torque structures arising from azimuthal pressure gradients, non-continuum terms in the Riemann problem, and feedback torque from SNe.

Within the disc, the structure of the pressure and non-continuum torques largely follow where SNe occur. Note, $\ell$=2 modes can still be seen in the iCGM in \textbf{m12m}, \textbf{m12i}, and \textbf{m12f}. These modes are offset from what is seen in the gravitational torques, as expected from gravitational torque theory of a collisionless component acting on a collisional component \citep{noguchi88,quataert11}. SNe torques tend to be dominant within the disc, while pressure torques become dominant in the iCGM. The torques arising from the non-continuum terms between gas cells lack the lower order structure of the pressure torques, with smaller scale variations. It is reasonable that they are largely limited to where SNe occur, as this is where unresolved hydrodynamic interactions will be present. Note, the pressure torques within the disc largely arise from thermal pressure gradients, while the pressure torques in the iCGM largely arise from CR pressure gradients (See Appx.~\ref{sec:appendix_misc}).

SNe torque manifests itself primarily as strong positive/negative torques around the sites of the SNe, with much more limited effects outside the disc. When looking at the torques from an individual star particle's SNe, the gas rotationally downstream (counter-clockwise) from the star particle has a positive torque, as the gas is being accelerated in the direction of rotation and is therefore being torqued up. Likewise, the gas rotationally upstream of the SNe is being pushed counter to the rotation and is therefore losing angular momentum. The asymmetry that arises from these torques will be discussed in more detail in Sec.~\ref{sec:disc_On-SNe}.

\subsection{Torques on Individual Mass Elements}

\begin{figure*}
	  \center{\includegraphics[width=.9 \textwidth]
 	       {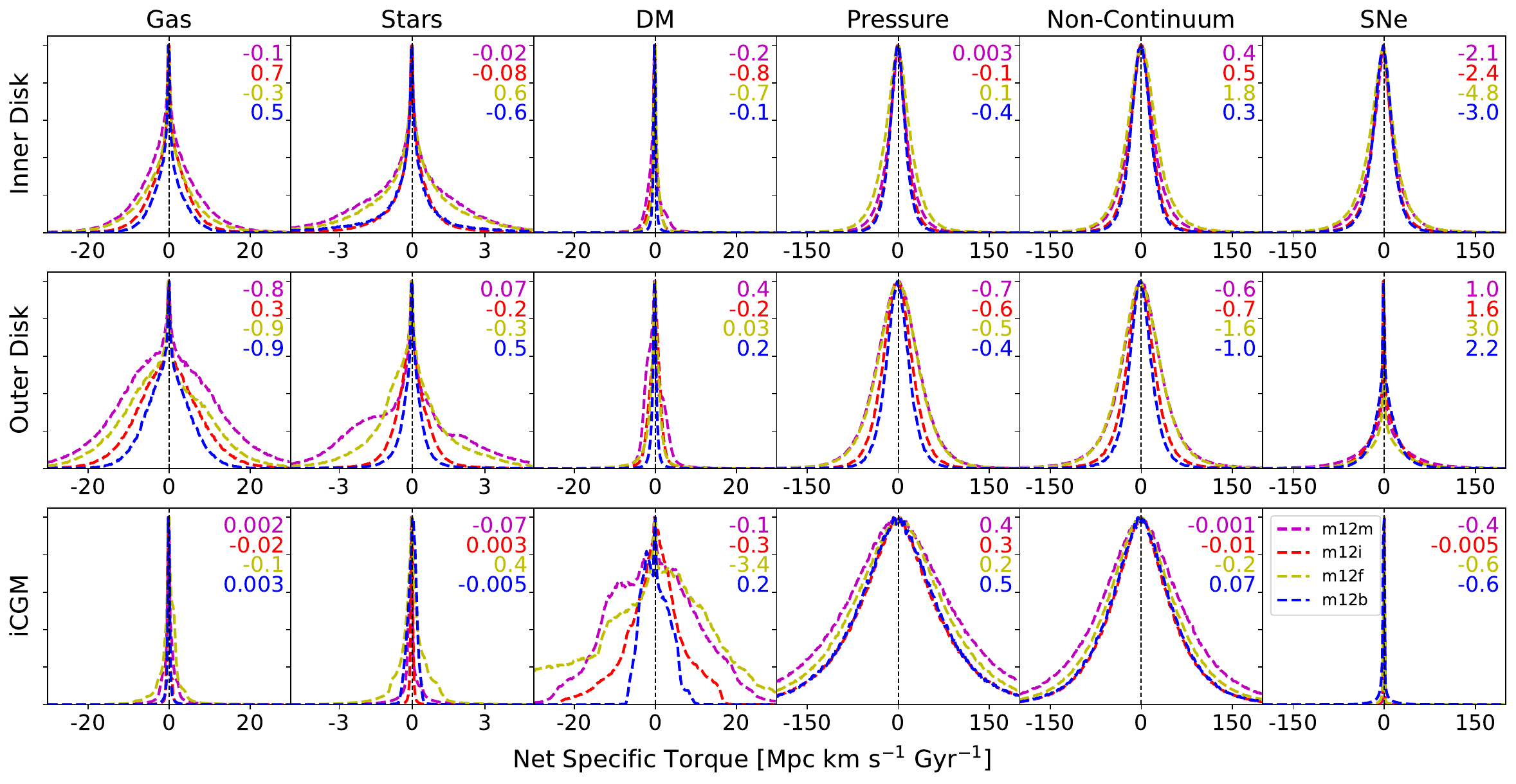}}
	  \caption{\label{fig:histPlotMain} Mass weighted histograms showing the PDF for the z-component of the net specific torques on gas mass elements over the 10 snapshots ($\sim$ 200 Myr at z=0) considered in this study. The values in the upper right corner of each plot show the mass-weighted means of the distribution. Mass elements are divided into three spatial regimes based on their average position over the 10 snapshots.  \textbf{Top Row:} inner disc; elements with a cylindrical radius within the inner 0.5 \rgal (and within 2 times the scale height of the gaseous disc. \textbf{Middle Row:} outer disc; elements with a cylindrical radius between 0.5 and 1 \rgal and within 2 times the scale height of the gaseous disc. \textbf{Bottom Row:} inner CGM, elements at spherical radial distances greater than 1 \rgal, but within 4 \rgal of the disc centre. Note, \rgal is the radius where the average hydrogen column density drops below 10$^{20.3} \rm{cm}^{-2}$ and is a measure of the size of the gaseous disc. We consider the gravitational torques from gas, stars, and dark matter, the hydrodynamical torques from the pressure and non-continuum terms in the Riemann problem, and the torque from SNe. Gravitational torques from gas and stars are more significant inside the disc than outside. The gravitational torque from the dark matter is more significant at larger radii, with the dark matter in the CGM providing a net inward torque in all cases. The hydrodynamical torques are a consistent inward torque within the outer disc. The torques from SNe are the most significant source of inward torque in the inner disc, and provide a smaller outward torque in the outer disc. The net contribution in the CGM is also inward, albeit to a lesser degree. Note the total gravitational torque on a given gas mass element was used to normalize the post-processing estimates from various sources at each snapshot to ensure the total gravitational torque was accurate.}
\end{figure*}

Throughout the accretion and radial transport process, individual gas mass elements may not experience the same net torque as azimuthal averages may imply. As they move, they pass through regions with different torques, allowing for individual mass elements to experience more or less inward torque than would seem apparent from the spatial averages as seen in other simulations, e.g. \cite{quataert11}. In Fig.~\ref{fig:histPlotMain}, we show the mass-weighted distribution of the dominant net torques (gravitational torques from gas, stars and dark matter, hydrodynamical torques from pressure gradients and non-continuum terms, and torques from SNe momentum injection) acting on the individual mass elements in three distinct spatial regions: the inner disc ($s\leq$ $\frac{1}{2}$\rgalStop), the outer disc ($\frac{1}{2}$\rgalStop $< s \leq$\rgalStop), and the iCGM (\rgalStop < $r$ $\leq$ 4 \rgal) \footnote{This upper limit for the iCGM was selected as it is a region where gas is still largely co-rotating with the disc and is close enough to feasibly interact with the disc in a short time period.} where $s$ is the gas mass element's average cylindrical radius over the 10 snapshots considered and $r$ is its average distance from the galactic centre. We additionally constrain the inner and outer disc regions to be within $\pm$2 \hgal~(Table~\ref{table:discs}) to limit our analysis to gas mass elements within the disc. The gravitational torques from gas, stars, and dark matter are normalized to the total gravitational torques.

Within the disc, the net contribution from gas and stellar gravity is at its strongest. Stellar torques provide a net negative torque in 3 galaxies in the inner disc, and 2 galaxies in the outer disc. The gravitational torques from the dark matter are significant everywhere and are the only significant gravitational torques in the iCGM, where they provide a net negative torque in all cases but \textbf{m12b}. The structure of the distribution of torques for gas and stars peaks sharply at zero, with larger wings within the disc. The dark matter torque structure is more complicated, with bumps at certain values depending on the galaxy.

The net hydrodynamical torques are of a similar order of magnitude as the gravitational torques within the disc and are more significant on average in the iCGM. The pressure torques provide a net negative torque in all cases in the outer disc and are less significant in the inner disc. Within the CGM, they provide a net positive torque in all cases. The torques arising from the non-continuum terms of the Riemann problem provide a net positive torque in the inner disc, a net negative torque in the outer disc, and are less significant within the iCGM. The distribution for both of these hydro torques is largely symmetric, with a much larger variance than the gravitational torques.

Finally, the net torques arising from SNe provide a strong negative torque within the inner disc and are largely dominant to the hydro and gravitational torques. In the outer disc, they are slightly smaller but provide a net positive torque. Within the iCGM, they are less significant and provide a small negative torque in most cases. The distribution for the SNe torques peaks at zero, with a noticeable asymmetry in the wings. It is worth noting that the asymmetries in the SNe net torques are opposite the non-continuum torques in all cases within the disc.

\section{Discussion} \label{sec:discussion}

The dominant source of torque on a given gas mass element depends strongly on its position in the disc as well as the timescale we are considering. Overall, we see that gas-gas gravitational interactions, pressure interactions, and supernovae effectively transfer angular momentum radially outwards. Gravitational torques from stars and dark matter act as a net sink of angular momentum in the innermost regions of the disc and the iCGM, respectively.

Gas and stellar gravitational torques are significant within the disc, but drop off as a function of radius. Averaged over time, the gas-gas gravitational interactions preferentially transfer angular momentum to just inside of the disc edge. The gravitational torques from the dark matter become most significant in the iCGM, with a distinct $\ell$=2 like spherical harmonic pattern in all cases. While accreting mass elements may experience both regions of positive and negative torque as they move through the dark matter halo, the net effect is a negative torque from dark matter in the iCGM. This implies the dark matter halo is acting as a sink for the angular momentum of the gas outside the disc. Note that the change in angular momentum in the iCGM is small and gas is not fully rotationally supported, so does not require additional torque to infall.

The hydro torques tend to be the most dominant when looking at a single time point, with values ranging up to 20 Mpc km s$^{-1}$ Gyr$^{-1}$ in the iCGM and within many smaller regions within the disc (Fig.~\ref{fig:hydroTorqueFacePlot}).
 Regions within the disc where the hydro torques tend to be the most significant are largely co-located with SNe. When considered over a full oscillatory period, the net effect of the hydro torques largely cancel out, contributing a smaller net negative torque. In the iCGM, the pressure terms are the dominant hydro term. Within the disc, they are similar to the non-continuum hydro terms, with the pressure terms being slightly more significant on average. We will discuss the effective angular momentum transfer of these gas-gas interactions in more detail in Section~\ref{sec:disc_L-exchange}.

The torques from stellar feedback also provide a surprisingly important source of torque within the disc.
The residual torques from SNe are of a similar order of magnitude to the hydro and gravity torques, depending on the spatial/temporal scales you are looking at. While the SNe torques are largely symmetric, there is still a net non-zero cancellation at a given radius. We will discuss the cause of this asymmetry in Section~\ref{sec:disc_On-SNe}.

\subsection{Time Dependence of Torques}\label{sec: disc_TimeDependence}

\begin{figure}
	  \center{\includegraphics[width=.5 \textwidth]
 	       {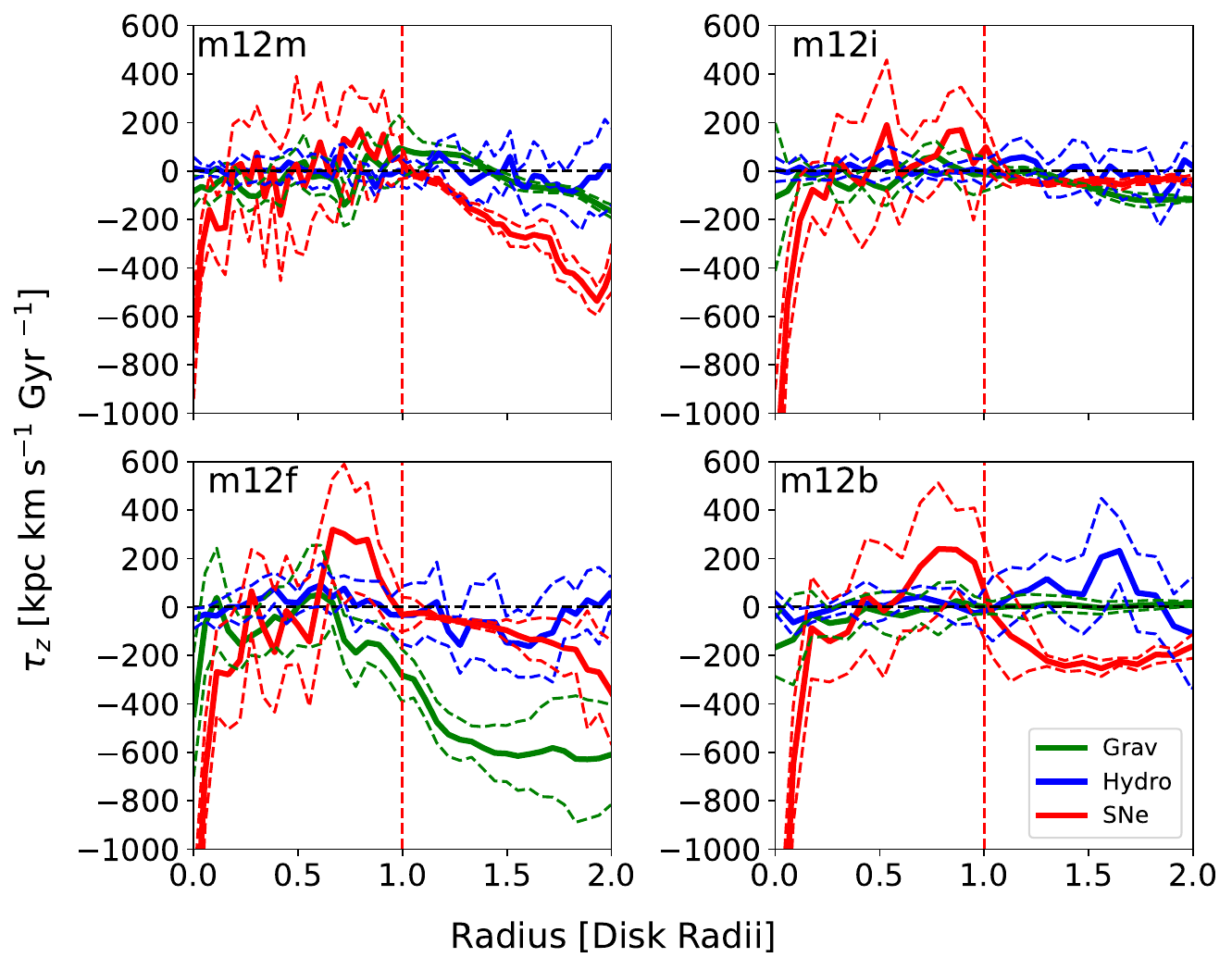}}
	  \caption{\label{fig:torqueStdv} Specific torque acting on gas from the three main sources with their corresponding standard deviations over time as a function of cylindrical radius normalized to the size of the gaseous disc. Values are considered between $\pm$10 kpc from the disc plane. The standard deviation in this case represents how the azimuthally averaged torques vary over time, not on an element-by-element basis.  Within the disc, the torques from SNe have the largest standard deviations, largely due to their more stochastic nature compared to the other sources of torque. The standard deviations of the hydrodynamical and gravitational torques are of similar order within the disc. In the innermost regions of the disc, where stellar gravitational torques become more important than gas-gas gravitational torques, the standard deviation increases. In the iCGM, where the dark matter gravitational torque is the dominant gravitational torque, the variance decreases significantly, and the variance from the hydrodynamical torques becomes the largest. Standard deviations are typically a few times larger than mean values within the disc. Except for the hydro torques, standard deviations are smaller than the mean values in the iCGM.
   }
\end{figure}

The quantitative plots presented in this study are all time and spatial averages, as there is a large amount of variance azimuthally and temporally over a dynamical time. While these averages represent the net angular momentum transfer through the system, in this section we will discuss how these terms vary over time.

Fig.~\ref{fig:torqueStdv} shows the temporal standard deviations between snapshots of the azimuthally averaged torques over time. In general, these temporal variances are smaller than the maximum element-by-element variances (Fig.~\ref{fig:histPlotMain}). Within the disc, these deviations are dominated by the torques from SNe. Correspondingly, looking at the projection maps of the SNe torques (Fig.~\ref{fig:hydroTorqueFacePlot}), we see that at a given snapshot there are a few star particles ($\sim$5-20) that dominate the SNe momentum injection into the surrounding gas at a given time. As time advances, the individual torque dipoles rotate with the disc and typically dissipate within $\sim$40 Myr. 
We will discuss the possibilities of how this large variance from a relatively small subselection of star particles leads to strong net torques in Sec.~\ref{sec:disc_On-SNe}.

The standard deviations of the gravitational torques vary based on whether they are dominated by stellar gravity, gas-gas interactions, or dark matter. The projection maps of the stellar gravitational torques largely rotate with the disc at larger radii, but the innermost radii are more chaotic leading to the higher standard deviations in the inner disc. Similarly, the large spatial patterns of the gas-gas gravitational torques rotate with the disc, but there are smaller-scale hotspots that vary between snapshots. The torque field from the dark matter does not evolve significantly with time, leading to low standard deviations in the iCGM.

The gas-gas hydrodynamical torques vary temporally at a similar level as the gas-gas gravitational torques within the disc and slightly more on average in the iCGM. The projection maps (Fig.~\ref{fig:hydroTorqueFacePlot},\ref{fig:pressureSubcomponentFaceplots}), show the temporal variations within the disc are largely dominated by the hotspots in the thermal pressure and non-continuum terms, and are largely correlated with the location of SNe. In the iCGM, the broad torque patterns are largely due to CR pressure torques and rotate with the gas. These patterns shift and vary on time scales of a few snapshots ($\sim$40-60 Myr), leading to the higher variations seen within the iCGM.

The total standard deviations within the disc are on the order of or several times larger than the mean values of the specific torques. Detecting the signatures of these flows in observations will therefore require large samples of galaxies for proper averages. We will discuss the secondary effects these torque structures may have on observations in more detail in Sec.~\ref{sec:observationalComps}.

\subsection{Angular Momentum Exchange between Gas} \label{sec:disc_L-exchange}

Torques between gas mass elements (gas-gas gravitational interactions and hydrodynamical torques) should cancel out when considering the system as a whole. When considering individual gas mass elements, however, we can identify which elements are gaining/losing angular momentum and infer where this angular momentum is ultimately being transferred. If gas mass elements are ejected from the disc through these interactions, it can lead to a net angular momentum loss.

The effects of gas-gas gravitational torques are important in the transport of angular momentum. On average, gas in the iCGM tends to lose a slight amount of angular momentum from these interactions, while gas just interior to the disc edge tends to gain angular momentum. Interactions within the disc are more varied, but there tends to be a net transfer of angular momentum towards the outer region of the disc.

Torques from the pressure gradient preferentially transfer angular momentum from the inner regions of the disc to the iCGM. Specifically, they deposit angular momentum just outside the gaseous disc edge. 
The thermal pressure contribution behaves similarly to the SNe torques, showing negative torques in the inner disc and peaking in the outer disc. The CR pressure torques are responsible for the peak just outside the disc edge and dominate the iCGM pressure torques (Fig.~\ref{fig:pressureSubcomponentRadialPlots}).

The torques arising from the non-continuum terms in the Riemann problem act in opposition to the pressure torques. 
These terms are expected to be significant in regions with many shocks, such as around areas of active star formation. While the magnitude of torques is typically less significant than those of the pressure gradients, they do reduce the overall transport of angular momentum out of the disc in the innermost radii.

The combined effect of the gas-gas gravitational torques and the pressure gradient torques transfer angular momentum from within the disc to the disc edge. This is of particular interest, as this is where gas is preferentially joining the disc \citep{trapp22}. While incoming gas from the iCGM loses a small amount of angular momentum as it transports near the disc, it gains angular momentum just prior to joining co-rotation and drops vertically downward onto the disc \citep{trapp22,hafen22,stern24:HotRotatingCGMInflows}. Compared with gas within the disc, this actively accreting gas is rotationally lagging and moving radially inwards at higher speeds on average. As this accreting gas hits the disc, it may be gaining angular momentum from gas already within the disc through these gas-gas interactions. This would simultaneously torque up the infalling gas, slow it down radially, and remove angular momentum from gas already inside the disc, allowing it to transfer inward. 

\subsection{On Supernovae Momentum Injection} \label{sec:disc_On-SNe}

\begin{figure}
	  \center{\includegraphics[width=.5 \textwidth]
 	       {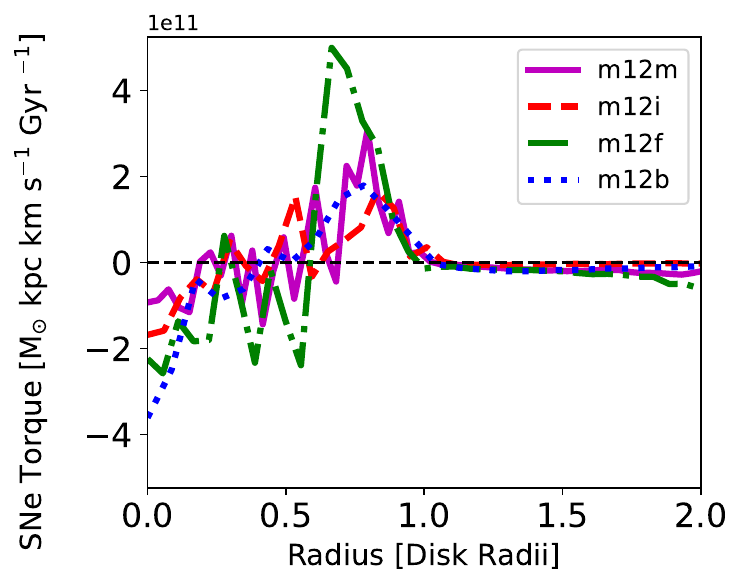}}
	  \caption{\label{fig:SneFullMom_RadialPlot}  The z-component of the total torque (as opposed to the specific torque) arising from SNe for the 4 galaxies of interest as a function of cylindrical radius within $\pm$10 kpc of the disc plane. Plots are averaged over 10 snapshots ($\sim$200 Myr) near redshift 0 and the radius has been normalized to the size of the disc edge as denoted by the HI column density transition. The horizontal black dashed line marks 0 torque. Negative values correspond to inflow. Torques are negative interior to the radius of co-rotation and become positive exterior to this radius (about 0.25-0.5 disc radii). Torques in the iCGM are very close to zero, except for \textbf{m12f}, which has a merger. When averaged over the whole disc, the SNe torques largely cancel out.
}
\end{figure}

The implementation of mechanical feedback from supernovae in the FIRE-2 simulations avoids imprinting "preferred directions" on the ejecta that may arise from anisotropic gas distributions surrounding the SNe. Specifically, effective faces for the gas mass elements that overlap the SNe event (either they lie within the star particle's smoothing length or vice versa) are constructed in a similar way as they are in the hydro solver. An effective weight for each face is calculated to ensure the conservation of energy and momentum. Special care is taken when determining the vector weights for the momentum, with each vector component receiving a separate weight. In effect, this accounts for the fact that the direction of the momentum injection vector may not be aligned with the normal vector of the effective face for an arbitrary anisotropic distribution of gas. See \cite{hopkins18b-SneMethods} for more details.\footnote{In FIRE-2, momentum injected by supernovae is initially fully isotropic. Additional momentum from the Sedov-Taylor phase is added via a sub-resolution solution based on local gas properties. As these solutions are applied independently in cones, there can be different amounts of $P dV$ work done in different directions \citep{hopkins18b-SneMethods}}

Figs.~\ref{fig:histPlotMain}, \ref{fig:torqueStdv}, and \ref{fig:SneFullMom_RadialPlot} show that while the instantaneous SNe torques are broadly both positive and negative (for gas rotationally downstream and upstream, respectively) and have a standard deviation much larger than their mean (as expected), there is a non-zero, statistically significant residual net feedback torque (unresolved torques from the sub-grid $PdV$ work done by SNe remnants in their expansion) that emerges after averaging spatially and temporally. These residuals are negative in the inner half of the disc and positive in the outer half of the disc. 
One might wonder whether this effect could arise purely from numerical discreteness effects in the initial cell coupling or cell positions/velocity updates, or from a numerical integration error in angular momentum over the salient timesteps. We address this in Appendix~\ref{sec:appendix_validationPlots}, where we show the upper limit to the total change in angular momentum from these effects is 1-3 orders of magnitude smaller than the net SNe torques shown in Fig.~\ref{fig:SneFullMom_RadialPlot}, and represent a sub-percent-level effect on the total net torque from all sources.

We hypothesize that the non-zero cancellation of the SNe torque we see at a given radius arises from the offsets between the spiral arm-like overdensities and the young stellar populations that generate the bulk of the SNe. These overdensities will preferentially be the sites of active star formation \citep{roberts69}, however depending on differences between the rotation curve and the spiral-pattern speed at a given radius, young stellar populations will either begin to lead or lag behind these overdensities. The stellar populations formed in these dense regions will begin producing supernovae explosions after a few Myr, with peak rates about a Myr later \citep{matzner02}. This is followed by a period of relatively consistent supernovae on the order of 30 Myr. Interior to the radius of co-rotation, wherein rotational speeds are faster than the spiral pattern speed, these young stellar populations will be rotationally downstream of the spiral overdensities by the time they enter peak supernovae rates. Such offsets have been seen in observations \citep{egusa09,louie13,egusa17-StellarSpiralOffsets-m51,schinnerer17-PAWS-SpiralArmsInCloudAndStarFormation,leroy17-ismStrucAndSF-m51,williams22-PhangsJwst-SpurringStarFormation,finn24-SubgalEnvironsOnMolCloud}. This ultimately may result in a slight asymmetry to the torques induced by these supernovae events, with the overdense regions being preferentially torqued inwards. At the radius of co-rotation, the pattern speed and the average rotational velocity will match, and we expect no net asymmetry in the sign of the supernovae torque. Past this radius, this trend will be inversed, with a net asymmetry towards positive torques, as the stellar populations will be moving more slowly than the pattern speed. This can be seen in Fig.~\ref{fig:SneFullMom_RadialPlot} in all cases. Effects from the drop-off in SFRs with radius, velocity asymmetries of ambient gas around expanding supernovae remnants, and SNe stochasticity may also play a role in modulating this signal. This would imply that, while these SNe torques are dominant through the disc, their net effect ultimately arises from the existing spiral structure. Similar to our discussion of the MHD torques, these SNe torques ultimately reinforce the preferred direction induced by the gravitational torques that initially form these structures. This is reflected in the net effects of the gas-gas gravitational, pressure, and SNe torques, which all effectively transport angular momentum radially outwards.

As discussed above, the weighting of the effective faces of the gas mass elements is done such that linear momentum is conserved. Focusing on the gas interior to the radius of co-rotation, gas in overdensities rotationally upstream from a SNe will experience negative torques and the gas downstream will experience positive torques to compensate. Given the spiral-like structure seen in the discs, the majority of this downstream gas may be located in overdensities at different radii. Since this gas is at different radii, it will not necessarily gain/lose the same amount of angular momentum, with gas at lower radii gaining less angular momentum for a given amount of momentum injected due to a shorter lever arm from the galactic centre. This allows for gas at a given radius to gain or lose angular momentum with respect to the galactic centre while linear momentum conservation of individual SNe is maintained. While this angular momentum is not guaranteed to be conserved in these events, there is a rough balance between positive and negative torques through the disc (Fig.~\ref{fig:SneFullMom_RadialPlot}).\footnote{Additionally, the small net angular momentum imparted from individual SNe does not reinforce this asymmetry (Fig.~\ref{fig:nonconservedTorque}).}

This picture is slightly more complicated in the iCGM, as there is little to no star formation and much less angular momentum transport. The net SNe torques will therefore be much more shot-noise dominated by the residuals of a few SNe. These net torques are negative in all cases and may arise from SNe near the disc edge or within nearby satellites. 
Note, \textbf{m12m} shows the highest specific torque in the CGM, and also has the most extended star formation. Similarly, \textbf{m12f} is undergoing a merger event, and shows larger SNe torques near 2 \rgalStop.

The significance of these feedback torques is not completely unexpected, as we are not the only study to see this. Using a completely different code, numerical methods, and feedback implementation, \cite{prieto16-multiscaleMassTransport} and \cite{prieto17-FeedbackEffectsOnMassTransportAtHighZ} found similar qualitative effects of pressure torques, likely driven by the “feedback torques”, on disc scales to what we describe here in the context of black hole fueling at z$\gtrsim$6.

\subsection{Comparison with previously characterized radial speeds}
\label{sec:disc_radialVelocities}

\begin{figure}
	  \center{\includegraphics[width=.45 \textwidth]
 	       {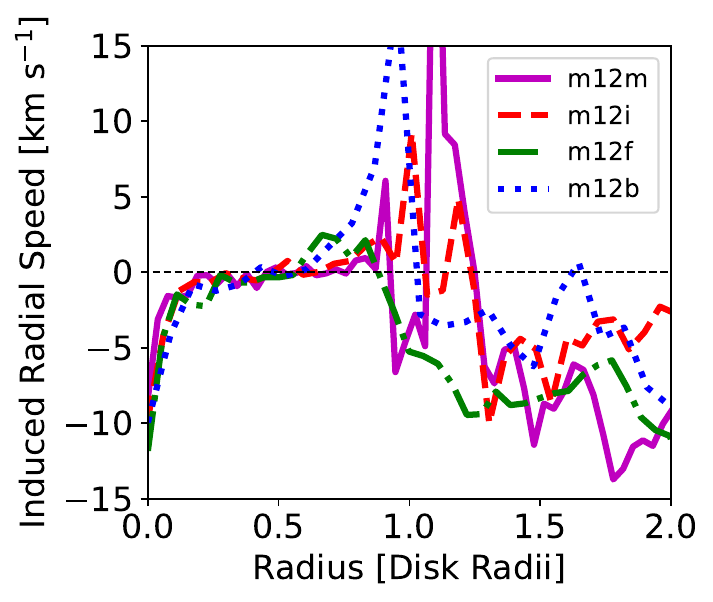}}
	  \caption{\label{fig:RadVelPredictions} The cylindrical radial speed the average measured torques would induce on the gas. Induced radial speed was calculated $\frac{dj/dt}{dj/ds}$, where $dj/dt$ is the average net torque and $dj/ds$ is the slope of the specific angular momentum curve as a function of cylindrical radius. In the inner half of the disc, torques induce radial speeds on the order of a few km s$^{-1}$, as expected from our first study \citep{trapp22}. In the centre-most regions, induced radial speeds are faster. The outer half of the disc shows positive induced radial speeds, corresponding to the slowdown of gas as it gains angular momentum and joins the disc. The nearby CGM has induced radial velocities on the order of 5-10 km s$^{-1}$. While less torque is needed to explain motion in this region, the slope of the specific angular momentum curve is also shallower, leading to larger predicted radial speeds.
}
\end{figure}

In \cite{trapp22}\footnote{\cite{trapp22} neglected the Hubble flow in these galaxies, which will lead to slightly smaller radial velocities in the outskirts by about 2-3 km/s at 2 \rgalStop and negligible changes in the inner disc. Overall conclusions are unaffected. Hubble velocity is included in this study where appropriate.} we quantified the radial velocities of gas onto and through these 4 galactic discs. Gas within the iCGM is moving towards the disc with average speeds of $\sim$10-20 km s$^{-1}$, as it is not fully rotationally supported. Gas gradually loses a small amount of angular momentum as it transports inwards through the CGM, and gains angular momentum just before joining the disc. Once within the disc, the average radial speeds of the gas slows down to $\sim$1-5 km s$^{-1}$. Gas in the disc is fully rotationally supported on average, implying the need for torque to explain this motion. While the dynamic structure of these galaxies is complex, in this section we will briefly compare the expected radial velocity values that would arise from the torques calculated in this study. Note, that caution is needed when interpreting changes in angular momentum as driving inflow/outflow, as this only applies if gas is in slowly decaying circular orbits and is homogenous and instantaneously well mixed. As shown in \cite{trapp22}, much of the gas is in quasi-circular orbits, so this should hold throughout most of the disc, but will not generally apply for gas that is already strongly inflowing/outflowing. For instance, large negative torques from a series of rapid-succession SNe, particularly in the galactic centre, may correspond to launching a parcel of gas out of the disc, either unbinding it or launching it into a highly radial orbit. This obviously does not correspond to the inflow of material.

When gas loses angular momentum, it will begin to sink radially inward. In the absence of any further interactions, it will oscillate around its new equilibrium radius, however, radial interactions and successive torques will complicate this trajectory.
In order to compare these torque values with what is to be expected by the average radial velocities, we assume that on average, as gas mass elements lose or gain angular momentum they follow the angular momentum curve (Fig.~\ref{fig:jPlot}). The average specific torque required to explain a given radial velocity would therefore be given by:
\begin{equation}
    \tau_{z} = v_{s} \frac{dj_{z}}{ds}
\end{equation}
where $\tau_{z}$ is the specific torque in the z-direction, $v_{s}$ is the cylindrical radial speed, and $\frac{dj_{z}}{ds}$ is the radial derivative of specific angular momentum in the z-direction.

Fig.~\ref{fig:RadVelPredictions} shows these induced radial speeds for the four galaxies. Within the inner disc, these torques will induce speeds on the order of a few km s$^{-1}$, as expected. In the most interior regions, this induces larger velocities primarily arising from torques from SNe and, to a lesser extent, stellar gravity. Gas in these inner regions may also turn into stars or get ejected. In the outer half of the disc, the induced radial speeds tend to be positive. Given that accreting gas preferentially joins the outer half of these discs, this likely corresponds to the radial slowdown of gas as it gains angular momentum and joins the disc. The iCGM has higher induced inward radial speeds on average, on the order of 5-10 km s$^{-1}$. While there is less overall angular momentum transfer in the iCGM, the slope of $dj_{z}/ds$ is also smaller, leading to larger predicted radial velocities. Given that gas in the iCGM is not rotationally supported, it likely has additional inward radial speed that would not be seen in this plot. This would account for why the induced radial speeds in the iCGM are slightly slower than measured in our previous study (10-20 km s$^{-1}$).

The largest instantaneous radial speeds we see in \cite{trapp22} are much higher ($\sim$40-60 km s$^{-1}$) than the average values. These speeds may arise from a variety of sources, including accelerations from radial pushes that impart no net torque. It is worth noting that if we extrapolate this analysis to higher velocities, the typical peak torque values seen in the face projections ($\sim$10,000 kpc km s$^{-1}$ Gyr$^{-1}$) would correspond to radial speeds around  ($\sim$50 km s$^{-1}$) assuming $dj_{z}/ds \sim$ 200 kpc Gyr$^{-1}$. Note, at a given time there will also be numerous radial forces and transient effects, so a one-to-one comparison of velocities and torques is complicated.

\subsection{Observational Implications}\label{sec:observationalComps}
The angular momentum transfer described in this study has implications for the dynamical and morphological structure of disc galaxies. As discussed in Sec.~\ref{sec: disc_TimeDependence}, there are large torques and radial forces present at a given time that can dominate the instantaneous velocity of gas, so some averaging (spatial and or temporal) is necessary to isolate the net effects.   
SNe in particular provide a strong negative torque in the innermost regions of all galaxies. This may lead to a build-up of density within the inner half of the disc. This will be limited by feedback effects, such as star formation-driven central outflows \citep{muratov15,roberts20-observedOutflows}. Based on the relative efficiency of the SNe torques versus outflows, there will be an equilibrium in the mass distribution profile. This may tie directly into the observation of exponential disc profiles \citep{freeman70}, which are likely influenced by the angular momentum transport within the disc. Additionally, this ties in with studies of AGN fueling, allowing for a significant amount of gas to be pushed to the inner few kpc \citep{alcazar21-hyperrefinement}. Further study of how this angular momentum and inflow structure relates to the outflow structure as a function of redshift is needed to make more concrete predictions.

SNe torques, as well as gas-gas gravitational and pressure torques, tend to transfer angular momentum to the outer regions of the disc. This should lead to gradually increasing disc sizes. As the typical angular momentum of the disc edge increases, there should be discrepancies between gas within the disc and gas joining from the nearby CGM. We see this in the larger discs (all but \textbf{m12b}) in this study (Fig.~\ref{fig:jPlot}). Given that gas actively joining moves largely parallel to the disc in the iCGM and drops vertically downward a few kpc inside the gaseous disc edge \citep{trapp22}, this may manifest as a stronger extraplanar rotational lag at these radii. We still see net inward radial velocities at these radii, despite positive average torques. This likely has to do with the accreting gas gaining angular momentum as it joins the disc and is torqued into full alignment, however, a deeper study of this interface could prove interesting.

Gas joins the disc preferentially in the outer half of the disc \citep{trapp22}, which is where torques are positive (Fig.~\ref{fig:totalTorqueLinePlot}). Gas interior to where it joins is pushed inwards, while more recently accreted gas in the outer half of the disc is pushed outwards, building up the disc. As predicted by inside-out growth, new gas layers accrete with more angular momentum, causing them to join at larger radii and continuing this process. The torques acting within the nearby CGM, while not needed to explain inflow rates, may have interesting implications for how this structure is built up, with pressure torques preferentially transferring angular momentum to the iCGM.

\section{Conclusion}

In this study, we analyze the angular momentum transfer in and around four low redshift L$_{*}$ disc galaxies simulated using the FIRE-2 model with feedback from cosmic rays. We specifically investigate the torques from gravitational forces, hydro forces, forces from radiative transfer, magnetic forces, and the feedback torque arising from SNe and stellar winds, which represents the unresolved MHD torques from expanding SNe blastwaves and superbubbles in the ISM. We further break down the gravitational terms into torques arising from gas, stars, and dark matter. We break the hydro terms down into torques arising from the pressure gradients, non-continuum terms, and viscous shearing. 

The key points of our paper are summarized below:

\begin{itemize}
  \item The dominant sources of torque acting on gas mass elements in these discs are those arising from gravitational forces, pressure gradients, and the sub-grid $P dV$ work done by SNe remnants interacting with gas on $\lesssim 10$ pc scales. The other sources of torque considered are subdominant.

  \item Gas within the iCGM is largely infalling, without the need for any torque to move onto the disc. There is still a small amount of angular momentum transfer at these radii, largely caused by gravity from dark matter and pressure gradients.

  \item Gravitational torques from dark matter are significant throughout the disc and are the only relevant gravitational torques within the iCGM. These torques show a distinct $\ell$=2 spherical harmonic pattern and average to negative torques within the iCGM. This can ultimately act as a sink for angular momentum, allowing for a net loss of angular momentum in the gas.

  \item Gravitational torques from stars are only relevant in the inner half of the gaseous disc, as that is where the majority of stellar mass is located. They are average to negative torques, but drop off sharply with radius. This can act as a sink for angular momentum from nearby gas mass elements, allowing for a net loss of angular momentum in the gas in the most interior regions of the disc.

  \item Gravitational torques from gas-gas interactions are more varied but are only significant within the disc itself. They preferentially transfer angular momentum outwards, just interior to the gaseous disc edge. This corresponds roughly with where accreting gas joins the disc.

  \item Torques arising from the pressure gradient provide a significant source of torque both inside the disc and the iCGM. These interactions end up transferring angular momentum radially outwards on average.

  \item The feedback torque from SNe provides an important source of torque within the disc. In the inner half of the disc, these torques average to negative values, providing a net source of angular momentum loss at these radii. In the outer half of the disc, these torques average to positive values. Outside the disc, these torques are much less significant.

  Ultimately, the torques from the gas-gas gravitational interactions, hydrodynamical, and SNe torques all reinforce one another, transferring angular momentum radially outward. Gravitational torques from dark matter and stellar structure can additionally remove angular momentum from the gas in the iCGM and innermost regions of the disc, respectively. This enables gas accreted in galactic outskirts to fuel star formation in the more central regions of disc galaxies over cosmological timescales.

  In future work, we plan on extending our analysis in the CGM, investigating in more detail the radial velocity and density structures that seem to be modulating the angular momentum transfer in these systems. We will also investigate how this picture changes as a function of redshift, investigating how these torques and radial flows behave in the transition from starburst galaxies to stable discs. The in-plane component of the torque vector (corresponding to the vertical components of forces) is also of particular interest both in how misaligned gas in the iCGM enters full alignment with the disc at late times, as well as how discs initially form and should be investigated in future work.

\end{itemize}

\section{Acknowledgements}
  
CT and DK were supported by NSF grant AST-2108324. Support for PFH was provided by NSF Research Grants 1911233, 20009234, 2108318, NSF CAREER grant 1455342, NASA grants 80NSSC18K0562, HST-AR-15800. CAFG was supported by NSF through grants AST-2108230, AST-2307327, and CAREER award AST-1652522; by NASA through grants 17-ATP17-0067 and 21-ATP21-0036; by STScI through grants HST-GO-16730.016-A and JWST-AR-03252.001-A; and by CXO through grant TM2-23005X. NM acknowledges the support of the Natural Sciences and Engineering Research Council of Canada (NSERC),  funding reference number RGPIN-2023-04901. Numerical calculations were run on the Caltech compute cluster “Wheeler,” allocations FTA-Hopkins/AST20016 supported by the NSF and TACC, and NASA HEC SMD-16-7592. The simulations presented here used computational resources granted by the  Extreme  Science and  Engineering  Discovery  Environment  (XSEDE),  which is supported by  National  Science Foundation grant no.  OCI-1053575, specifically allocation TG-AST120025 and resources provided by PRAC NSF.1713353 supported by the NSF; Frontera allocations AST21010 and AST20016, supported by the NSF and TACC; Triton Shared Computing Cluster (TSCC) at the San Diego Supercomputer Center. The
data used in this work were, in part, hosted on facilities supported by
the Scientific Computing Core at the Flatiron Institute, a division of
the Simons Foundation. This work was performed in part at the Aspen Center for Physics, which is supported by National Science Foundation grant PHY-2210452. We would additionally like to thank the anonymous reviewer, whose insightful comments markedly improved the manuscript.

This research also made use of MATPLOTLIB \citep{matplotlib}, NUMPY \citep{numpy}, SCIPY \citep{scipy}, H5PY \citep{h5py}, pytreegrav (https://github.com/mikegrudic/pytreegrav), Meshoid (https://github.com/mikegrudic/meshoid), and NASA’s Astrophysics Data System.

\section{Data Availability}
The data supporting the plots within this article are available on reasonable request to the corresponding author. A public version of the GIZMO code is available at http://www.tapir.caltech.edu/~phopkins/Site/GIZMO.html.

Additional data including simulation snapshots, initial conditions, and derived data products are available at http://fire.northwestern.edu.

\bibliography{mybib}


\appendix

\section{Additional Torques}\label{sec:appendix_misc}

\begin{figure}
	  \center{\includegraphics[width=.5 \textwidth]
 	       {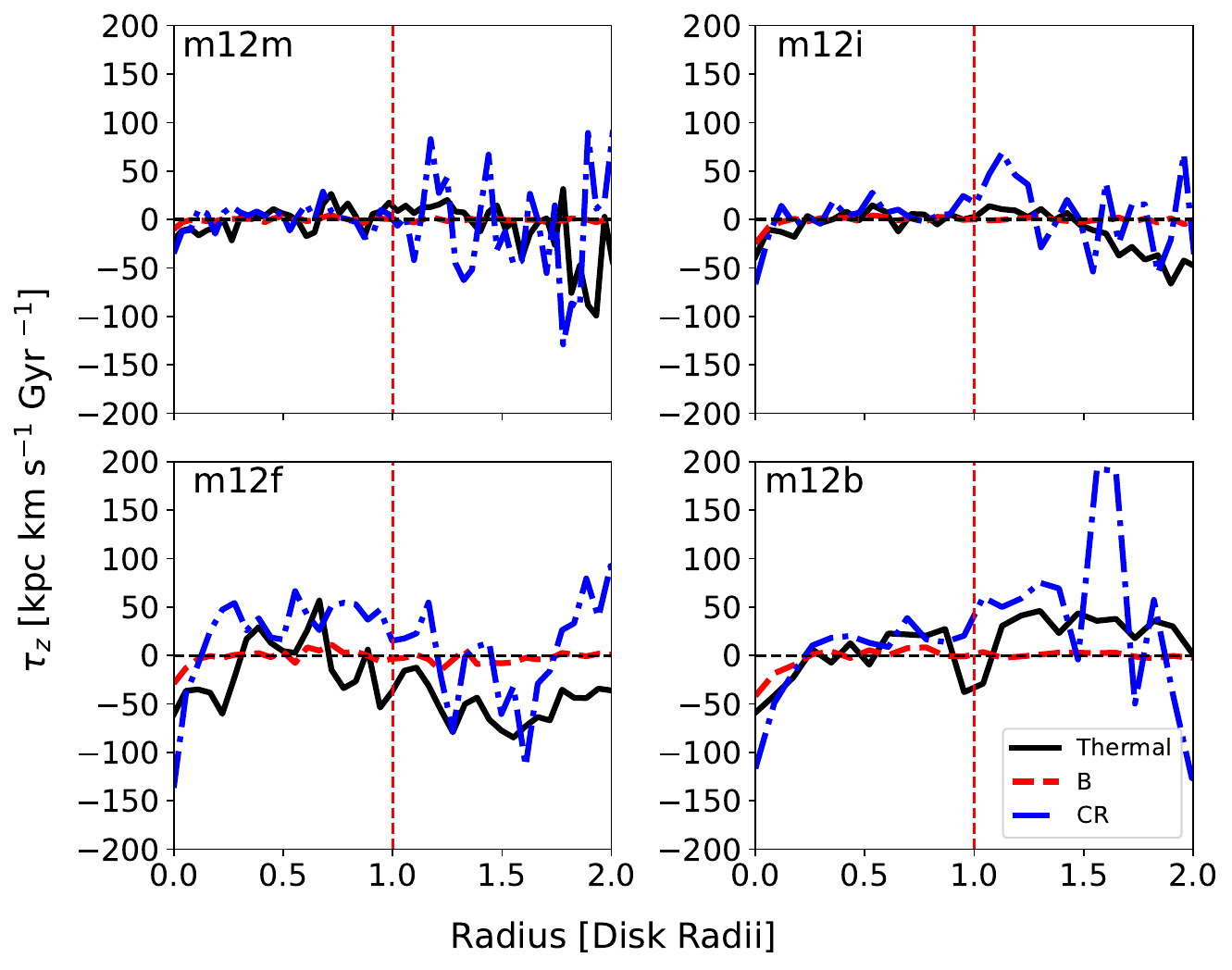}}
	  \caption{\label{fig:pressureSubcomponentRadialPlots} Azimuthally averaged mass-weighted specific torques for the 3 pressure subcomponents: thermal, CR, and magnetic. Thermal and CR pressure torques tend to be similarly important within the disc, while CR pressure torques dominate in the iCGM. The shape of the thermal pressure torque curve is similar to the SNE torques within the disc, showing negative torques in the inner half of the disc and positive torques in the outer half of the disc. The CR pressure torques tend to peak in the iCGm and are largely responsible for the peak just outside the disc edge present in most galaxies.
}
\end{figure}

\begin{figure}
	  \center{\includegraphics[width=.5 \textwidth]
 	       {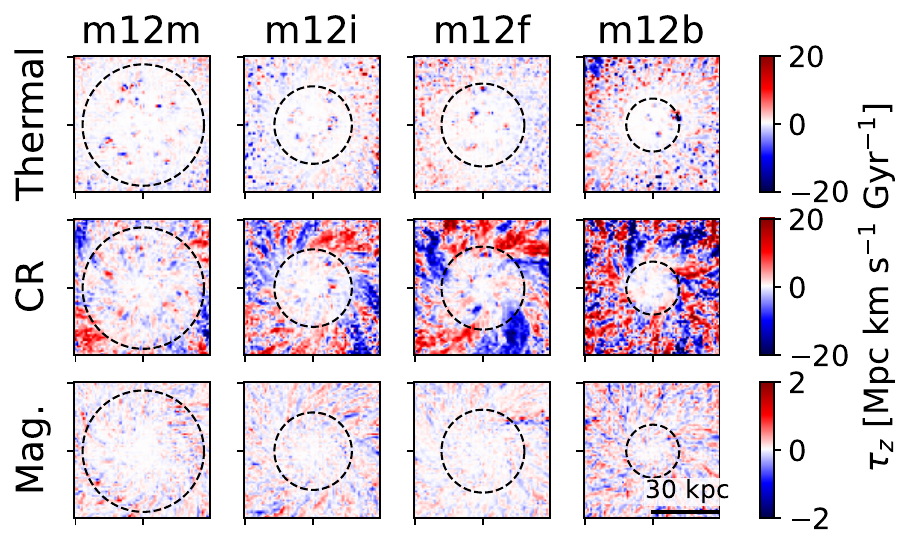}}
	  \caption{\label{fig:pressureSubcomponentFaceplots} Face-on plots showing mass-weighted average specific torques for the 3 pressure subcomponents: thermal pressure, CR pressure, and magnetic pressure. Positive torques (red) roughly correspond to outward motion, while negative torques (blue) roughly correspond to inward motion. The top row shows the total pressure torque measured directly from the simulation. Thermal pressure tends to be the most important within the disc, while CR pressure tends to be more significant in the iCGM. The magnetic pressure is sub-dominant by an order of magnitude.
}
\end{figure}

We additionally characterize the individual contributions from pressure subcomponents as calculated during the simulation run. In Fig.~\ref{fig:pressureSubcomponentRadialPlots}  we show the time-averaged torques as a function of disc radius for the thermal pressure, CR pressure, and magnetic pressure. The corresponding projection maps are shown in Fig.~\ref{fig:pressureSubcomponentFaceplots}. The torques from the thermal pressure have a similar structure as the SNe, peaking at hot spots throughout the disc with a characteristic structure. They behave the same quantitatively as well, with negative torques in the inner half of the disc, positive torques in the outer half of the disc, and a tendency for negative torques in the iCGM in most cases. The torques from the CR pressure are similarly significant within the disc, and become the dominant pressure term in the iCGM. They are responsible for the peak just outside the disc edge.  The torques from magnetic pressure are subdominant by an order of magnitude.

For completeness, we characterize the less significant sources of torque acting on the gas mass elements in this study, including torques from viscous forces, magnetic tension, stellar winds, and radiative transport. The magnetic tension term shown in these plots includes the divergence cleaning term. These torques are calculated in the same two ways as described in Sec.~\ref{sec:methodology}, although torques from the stellar wind terms could only be calculated following Eq.~\ref{eq:2}.

\begin{figure*}
	  \center{\includegraphics[width=.75 \textwidth]
 	       {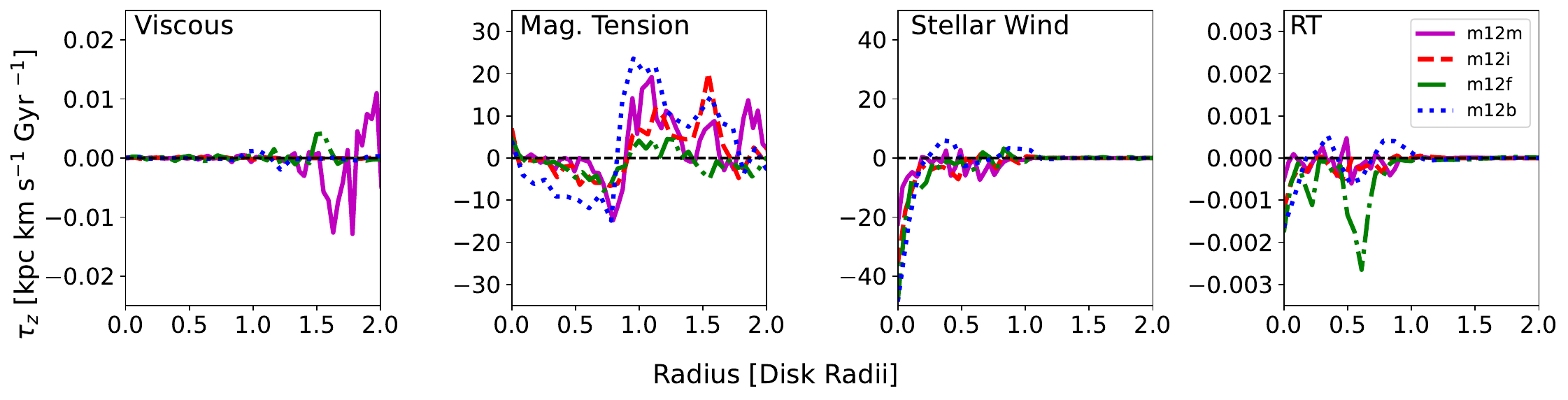}}
	  \caption{\label{fig:miscTorqueLinePlot}  Subdominant torques averaged across 10 snapshots ($\sim$ 200 Myr). Torques from viscous shearing (\textbf{left}) are irrelevant near the disc, but become more significant further out in the CGM as gas becomes increasingly misaligned with disc rotation. Torques from magnetic tension (\textbf{centre, left}) peak at disc edges. Torques from Stellar Winds (\textbf{centre, right}) and Radiative Transfer (RT, \textbf{right}) do show net negative torques.
}
\end{figure*}

In Fig.~\ref{fig:miscTorqueLinePlot} we show the averaged torques as a function of disc radius. Viscous torques become increasingly significant at higher radii but are largely subdominant. Torques arising from magnetic tension are most significant at the disc edge. Torques from stellar winds show negative torques on average in the inner disc but become much less significant at higher radii, where there is less star formation. Torques from radiative transfer are the least significant.

\begin{figure}
	  \center{\includegraphics[width=.5 \textwidth]
 	       {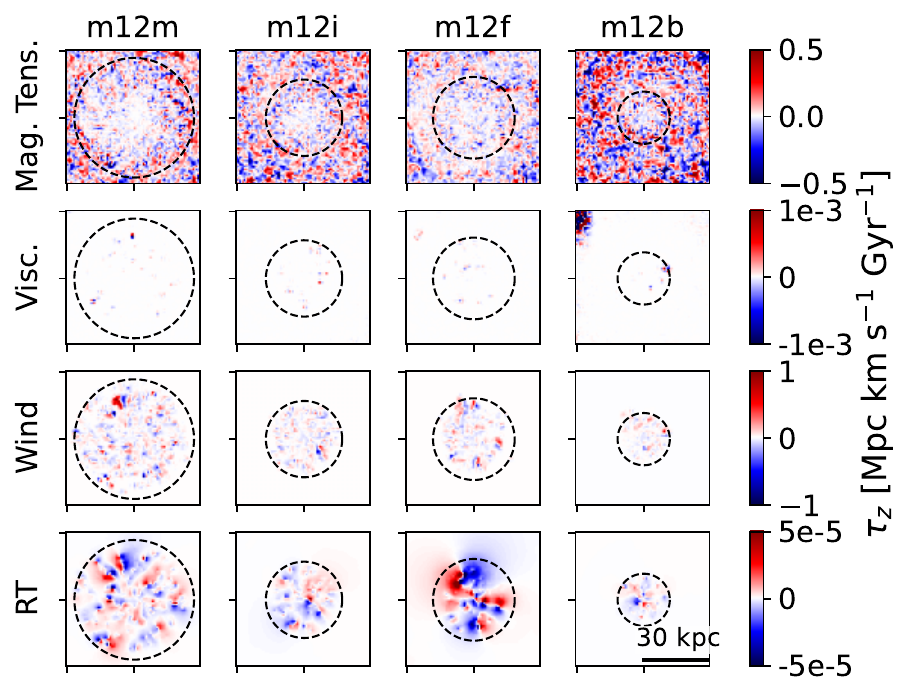}}
	  \caption{\label{fig:miscTorqueFacePlot} Face-on plots showing mass-weighted average specific torques for the 4 subdominant torque sources. Positive torques (red) roughly correspond to outflow, while negative torques (blue) roughly correspond to inflow. The top row shows the magnetic tension torques. The top-middle row shows the viscous torques, which become more relevant at larger radii but are still insignificant at these scales. The bottom two rows show the torque from stellar winds and radiative transfer, respectively. In both cases, the structure around stellar populations is similar to the SNe momentum injection torque structure (see Fig.~\ref{fig:hydroTorqueFacePlot}), but is quantitatively much less significant.
}
\end{figure}

In Fig.~\ref{fig:miscTorqueFacePlot} we show the corresponding projection plots for these torques. The magnetic tension torques have a relatively high-order structure, with spatial scales on the order of $\sim$1 kpc and a distinct structure at the disc edge. This can be seen most clearly in \textbf{m12m} and likely owing to B-field turbulent structure in our simulations \citep{ji21:virial_shocks_in_cr_halos}. The viscous torques have a similarly high-order structure, growing more significant further in the CGM, where gas is less well aligned with disc rotation and viscous shearing becomes increasingly important. The torques from stellar winds have a very similar structure as the torques from SNe, which is expected as they both mostly arise from young stellar populations. They have a much smaller extent than the SNe, however, especially in the CGM. The RT torques show a similar $\ell$=2 structure as the stellar gravitational torques, although offset and at a much lower order of magnitude. Similar structures to the SNe and stellar wind torques can also be seen, most clearly in \textbf{m12m}.

\section{Validation Plots}\label{sec:appendix_validationPlots}

\begin{figure}
    \center{\includegraphics[width=0.5 \textwidth]
 	       {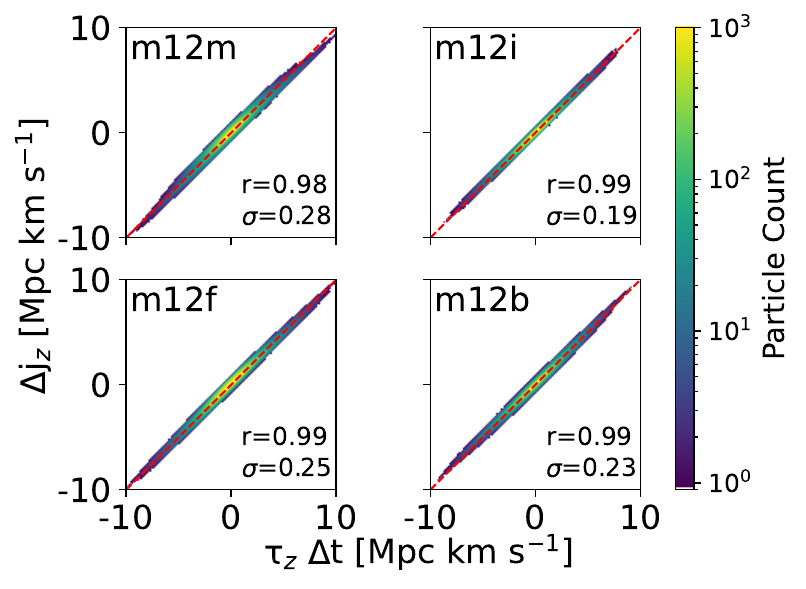}}
	  \caption{\label{fig:djdtScatter} Figure comparing the change in specific angular momentum of each gas mass element across the 10 snapshots considered as determined by that mass element's velocity (y-axis) ($j_{z,\rm{Final}}$-$j_{z, \rm{init}}$), versus the change in specific angular momentum from the sum of the considered torques (x-axis). Almost all gas mass elements match up as expected, with a characteristic variance. Slight differences likely arise from small errors in our estimate of the galactic centre (Sec.~\ref{sec:methodology}). To assess the tightness of this relation, we calculate the linear correlation (r) and scatter ($\sigma$). Correlations are very close to 1 as expected, with scatters around 0.2-0.3 Mpc km s$^{-1}$ (p$\ll$0.0001 in all cases, as expected from such a large sample size). 
      }
\end{figure}

Calculating the torques acting on each gas mass element is very sensitive to a variety of factors. In this implementation, the primary concern will be the centring of the galaxy in between snapshots, which is used when calculating the changes in angular momentum arising from each force. To verify we are reconstructing the total change in angular momentum, we plot the total calculated torque value multiplied by the snapshot time-step versus the change in specific angular momentum as measured by the gas mass elements' velocity vectors between two snapshots in Fig.~\ref{fig:djdtScatter}. The distribution is clearly linearly related with a small scatter (r$\sim$1; $\sigma\sim$0.2-0.3 Mpc km s$^{-1}$), implying we are reliably recreating the changes in angular momentum. Note that the color scale is shown in log scale to be able to visualize the scatter.

\begin{figure}
    \center{\includegraphics[width=0.5 \textwidth]
 	       {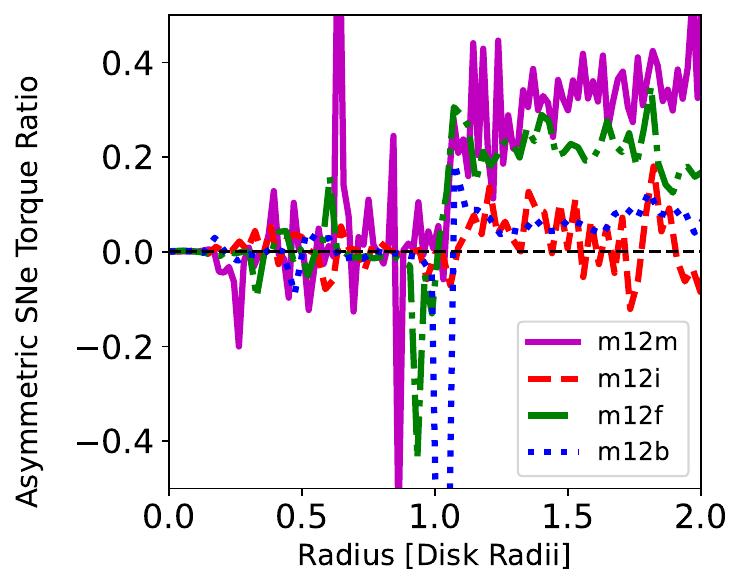}}
	  \caption{\label{fig:nonconservedTorque} Torques from deviations in isotropy of SNe from star particles divided by the net SNe torque acting on gas at a given cylindrical radius. For a given supernova event, the torque vector acting on all affected gas mass elements was summed together. Nonzero torque sums were then summed radially within $\pm$10 kpc of the disc plane, as in Fig.~\ref{fig:SneFullMom_RadialPlot}. Within the disc, this anisotropic contribution is at least two orders of magnitude smaller than the net torque acting on gas, aside from spikes where torques are near zero at half the disc radius. This implies that the SNe torques we see in the disc are not a consequence of numerical effects. Within the iCGM, this ratio is higher. This is due to two factors.  Firstly, the level of net torques from individual SNe is slightly higher, likely due to resolution limitations. Secondly, the net SNe torque acting on gas elements is drastically smaller, leading to a higher overall ratio. The ratio of total net torque from individual SNe to total torque acting on the gas mass elements is 0.004 in the inner disc and 0.01 in the outer disc.
      }
\end{figure}

\begin{figure}
    \center{\includegraphics[width=0.5 \textwidth]
 	       {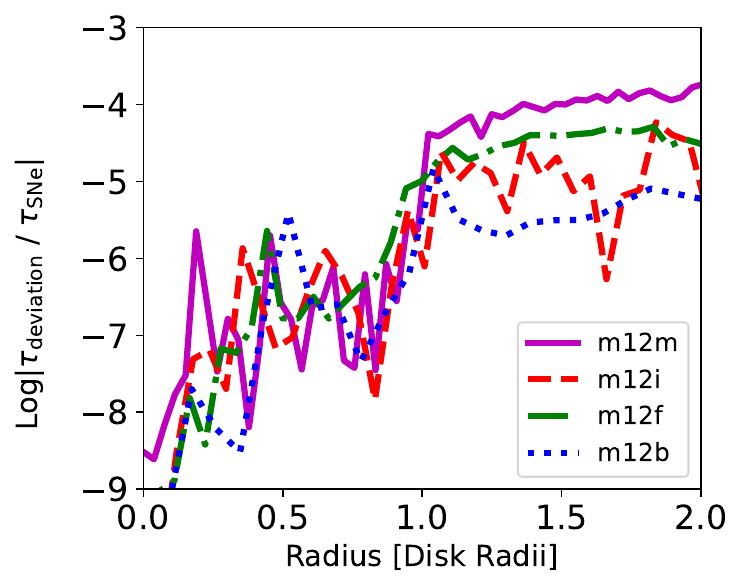}}
	  \caption{\label{fig:nonconservedTorqueFromLinearMomentum} 
   Same as Fig.~\ref{fig:nonconservedTorque}, but calculating the deviations from isotropic SNe torques by using the anisotropic linear momentum imparted by each SNe a given radius to calculate the torque based on the star particle's position, divided by the net torque the gas is experiencing at that radius. Note, the y-axis is in log scale and this ratio is several orders of magnitude smaller than that shown in Fig.~\ref{fig:nonconservedTorque}. Small deviations from isotropy in linear momentum input from the FIRE SNe algorithm have a negligible effect on net torques from the SNe.
   }
\end{figure}

Another concern is the effect of anisotropic momentum from SNe which leads to a sub-resolution generation of angular momentum. To address this, we compare the net feedback torque the star particles at a given radius impart on the system versus the net feedback torques acting on the gas at a given radius. This essentially gives a metric of how significantly numerical effects may be biasing our results. In Fig.~\ref{fig:nonconservedTorque}, we show that this net torque from the star particles is 1-3 orders of magnitude lower than the torque acting on the gas within the disc. This ratio is also not strictly positive or negative within the disc and does not follow the same trend seen in the feedback torque, implying it is not systematically biasing our conclusions. Within the iCGM, this ratio is higher on average. This is due to both slightly higher residual torques, as well as much lower overall torques due to the relatively small number of SNe in the iCGM. The spikes seen, particularly in \textbf{m12m} are largely due to the SNe torque values being near zero near half a disc radii. The ratio of total torques from deviations in isotropy of SNe from star particles to total torque acting on the gas mass elements is 0.004 in the inner disc and 0.01 in the outer disc. We additionally present a separate test in Fig.~\ref{fig:nonconservedTorqueFromLinearMomentum}, which instead of directly using the anisotropic angular momentum imparted onto the system by each SNe, uses the anisotropic linear momentum to calculate the associated torque at that star particles given position. At all radii, this ratio is several orders of magnitude smaller than the previous test.

\end{document}